\documentclass{article}
\usepackage{emulateapj}
\usepackage{graphicx}
\usepackage{psfig}
\usepackage{onecolfloat}
\submitted{accepted for publication in ApJS}
\newcommand\hfso{$^{\rm{o}}$}
\newcommand\etal{et al.}

\newcommand\gf{$gf$}

\newcommand\teff{T$_{\rm eff}$}
\newcommand\logg{log {\it g} }

\newcommand\simgt{\lower.3ex\hbox{\gtsima}}
\newcommand\mlogg{{\rm log}g}
\newcommand\Pd{\partial}
\newcommand\ep{\rm{log}\epsilon}
\newcommand\ltsim{$\scriptscriptstyle \; \buildrel < \over \sim \;$}
\begin{document}
\twocolumn [
\title{Abundances of 30 elements in 23 metal-poor stars}
\author{Jennifer A. Johnson}
\affil{OCIW, 813 Santa Barbara St., Pasadena, CA 91101}
\begin{abstract}We report the abundances of 30 elements in 23 metal-poor
($\lbrack$Fe/H$\rbrack$ $<-1.7$) giants.  These are based on 7774 equivalent widths
 and spectral synthesis
of 229 additional lines.  Hyperfine splitting is taken into account when 
appropriate.  Our choice of model atmospheres has the most influence 
on the accuracy of our abundances.  We consider
the effect of different model atmospheres on our results.  In addition
to the random errors in \teff, log {\it g}, and microturbulent velocity, there
are several sources of systematic error.  These include using \teff{} 
determined from \ion{Fe}{1} lines rather than colors, ignoring NLTE effects on
the \ion{Fe}{1}/\ion{Fe}{2} ionization balance, using models with solar 
$\lbrack \alpha$/Fe$\rbrack$
ratios and using Kurucz models with overshooting.  Of these, only the
use of models with solar $\lbrack \alpha$/Fe$\rbrack$ ratios had a negligible effect.
However, while the absolute
abundances can change by $>$ 0.10 dex, the relative abundances, especially
between closely allied atoms such as the rare earth group, often show
only small ($<$0.03 dex) changes.  
We found that some strong lines of
\ion{Fe}{1}, \ion{Mn}{1} and \ion{Cr}{1} consistently gave lower abundances by $\sim$0.2 dex,
a number larger than the quoted errors in the \gf{} values.  After
considering a model with depth-dependent microturbulent velocity and
a model with hotter temperatures in the upper layers, we conclude
that the latter did a better job of resolving the problem and agreeing
with observational evidence for the structure of stars.  The error analysis
includes the effects of correlation of \teff, log {\it g}, and $\xi$ errors, which
is crucial for certain element ratios, such as $\lbrack$Mg/Fe$\rbrack$.
The abundances
presented here are being analyzed and discussed in a 
separate series of papers.

\end{abstract}

\keywords{stars:abundances --- stars: atmospheres}
]

\section{Introduction}
Abundance ratios in 
metal-poor stars show the earliest stages of Galactic
chemical evolution.  These stars were polluted by metal-poor Type II SNe,
which have a different structure and nucleosynthesis from their
present-day metal-rich counterparts (e.g. Maeder 1992; 
Woosley \& Weaver 1995).  Fewer  
Type II SNe
have contributed to the abundances in a metal-poor star than to a star
with solar metallicity, so it is possible
to study the yields of individual SNe.  These facts make abundance
ratios in metal-poor stars very informative.
The survey by Beers, Preston, \& Shectman (1992) has expanded the number of 
known stars with [Fe/H]\footnote{We use the
usual notation [A/B]$\equiv \rm{log}_{10}
(N_A/N_B)_*-\rm{log}_{10}(N_A/N_B)_\odot$ and log$\epsilon(\rm A)\equiv \rm{log}_{10}(N_A/N_H)+12.0$.}$< -3.0$ by a factor of seven.  The more recent
Hamburg/ESO survey (Christlieb \& Beers 2000) is even more effective at
finding stars with [Fe/H] $< -2.0$, with 80\% of its candidates shown to 
be metal-poor stars.
Subsequent follow-up of metal-poor candidates from these surveys 
with high-resolution echelle data, particularly  
by McWilliam \etal{} (1995b) and Ryan, Norris \& Beers (1996),
showed two previously unobserved phenomena.  First, below
[Fe/H]$\sim -2.5$, [Mn/Fe] and [Cr/Fe] decrease with decreasing
metallicity, while [Co/Fe] increases.  These elements were
also tightly correlated among themselves.  The [Co/Cr] values changed
by $\sim$ 1 dex between [Fe/H]=$-$4.0 and $-$2.0, with very little dispersion.
Second, the dispersion in
abundance ratios is much more marked.  Previous investigations (e.g. Gilroy \etal{} 1988)
had supported a dispersion in the neutron-capture elements at
a given iron abundance; new observations showed that dispersion
in [Sr/Fe], for example, can be up to 2 dex (see also Depagne \etal{} 2000).  
These new studies took advantage of the large wavelength coverage of modern
echelles + CCDs to measure the abundances of many elements in one star,
which allow them to find these new correlations. 
In this paper, we expand the sample of 
metal-poor stars with many elements measured.  Our sample concentrates
on the brighter giants from the survey of Bond (1980), 
and is skewed toward somewhat more metal-rich
stars ($-3.05 < \rm{[Fe/H]} < -1.7$) than 
McWilliam \etal{} (1995) and Ryan \etal{} (1996).  About half of
our stars have [Fe/H] $< -2.5$, in the region of interest for
the iron peak elements, and all but three have [Fe/H] $< -2.0$, the region
where the dispersion in the neutron-capture elements is greatest.
We report the abundances of 30 elements in 22 metal-poor
field giants and 1 M92 giant.  In \S 2, we review our observations 
and data reduction. \S 3 discusses our choices for model atmospheres.
Since abundance ratios can be very sensitive to choices of model
atmospheric parameters, we discuss several potential problems.
We examine the effect on the abundances if we made different assumptions
when selecting model atmosphere parameters.  We used \teff{} derived using
spectroscopic data, rather than colors.  We did not consider NLTE effects
on \ion{Fe}{1} when adopting a \logg{}.  We also used Kurucz models which had
solar [$\alpha$/Fe] ratios.  Both the use of MARCS models and of $\alpha$-enhanced models could affect our results, as could the treatment of convection
and the temperature structure in the upper layers of the Kurucz model 
atmospheres.  We discuss each of these cases and show the resulting errors in 
the abundances.  Our final abundances are presented in \S 4.
These abundances are being analyzed and discussed in a separate series
of papers (Johnson \& Bolte 2001).  

\section{Observations and Data Reduction}

\subsection{Observations}

\begin{figure}[htb]
\begin{center}
\includegraphics[width=3.0in,angle=0]{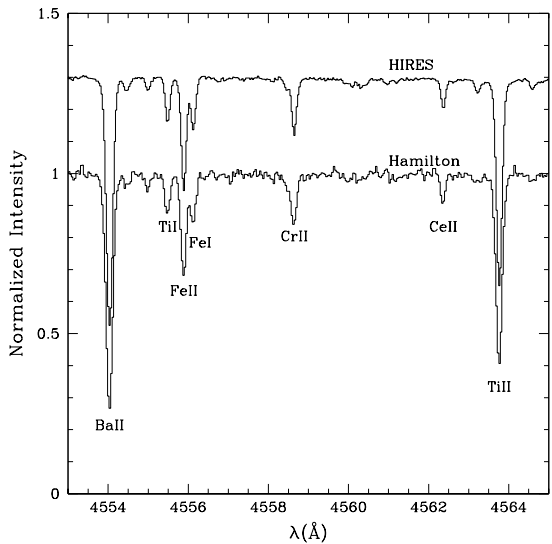}
\caption{Examples of HIRES (top) and Hamilton (bottom) data for
HD 186478 in a region of the spectrum where they overlap.}
\end{center}
\end{figure}

Metal-poor field giants from the lists of Bond (1980) 
were observed with HIRES on Keck I (Vogt \etal{} 1994) and the Hamilton spectrograph on
the Lick Shane 3-meter telescope (Vogt 1987).  The HIRES spectra cover 3200-4700
\AA{} with $R \sim 45,000$.  We used the C1 Decker and
 a 0.86$\arcsec$ slit.  The Hamilton spectra ($R \sim 60,000$)
cover almost the entire optical wavelength range, but useful data were
generally obtained between 4000-7100 \AA.  The stars were observed
through a 1.1$\arcsec$ slit.  Figure 1 shows examples of a HIRES and
a Hamilton spectrum in the wavelength range $\lambda \lambda$ 4553-4565 \AA.
We preferentially selected  
stars with [Fe/H] $< -2.0$ that either did not have high S/N
data with large wavelength coverage reported in the literature, 
or had known super-solar ratios of [Eu/Fe].  Eu is a neutron-capture
element, and a large enhancement of Eu would
probably allow us to measure the
abundances of many other neutron-capture elements.  We also
observed one star in M92 that Shetrone (1996) had shown to be
have [Eu/Fe]$\sim$0.4.  Table 1 lists the stars observed, the date of
observation, the S/N obtained, and the total integration time.  For
calibration, we took quartz-lamp flat fields, Th-Ar lamps, and
zero-second bias frames.  The HIRES data were binned by two in the
spatial direction.  The Hamilton data have very few pixels between
orders, so they were not binned.

\begin{deluxetable}{lccccc}
\small
\tablewidth{0pt}
\tablecaption{Observations of Metal-Poor Stars}
\tablenum{1}
\tablehead{\colhead {Star} & \colhead{$V_{\rm mag}$} & 
\colhead{UT Date} & \colhead{Instrument}
 & \colhead{Total Exposure} & \colhead{S/N at} \\
 & \colhead{Bond 1980} & & & \colhead{Time (seconds)}
 & \colhead{4700 \AA}}
\startdata
HD 29574  & \phn 8.38 & 9 March 1999 & Hamilton & 1800 & \phn 75  \\ 
HD 63791 &  \phn 7.90 & 7 March 1999 & Hamilton & 2700 & 100 \\ 
HD 88609 &  \phn 8.61 & 12 May 1997 & HIRES & 1200 & 350 \\
 & & 9 March 1999 & Hamilton & 2700 & 110 \\
HD 108577 & \phn 9.58 & 12 May 1997 & HIRES & 1800 & \\
 & & 20 June 1997 & HIRES & 1800 & \raisebox{1.5ex}[0pt]{400} \\
 & & 7 March 1999 & Hamilton & 3600 & 65 \\ 
HD 115444 & \phn 8.98 & 12 May 1997 & HIRES & 2400 & \\
 & & 20 June 1997 & HIRES & 1200  & \raisebox{1.5ex}[0pt]{400} \\
 & & 8 March 1999 & Hamilton & 3600  & 90 \\ 
HD 122563 & \phn 6.21 & 12 May 1997 & HIRES & \phn 600  & \\
 & & 20 June 1997 & HIRES & \phn 600 & \raisebox{1.5ex}[0pt]{500} \\
 & & 9 June 1998 & Hamilton & \phn 120 & 95 \\
HD 126587 & \phn 9.12 & 12 May 1997 & HIRES & 1800 & 350\\
 & & 25 May 1999 & Hamilton & 1800 & 50 \\
HD 128279 & \phn 8.04 & 12 May 1997 & HIRES & \phn 600 & 200 \\
 & & 8 March 1999 & Hamilton & 1800 & 30 \\ 
HD 165195 & \phn 7.34 & 11 August 1998 & Hamilton & \phn 900 & 100 \\
HD 186478 & \phn 9.16 & 12 May 1997 & HIRES & 1800 & \\
 & & 20 June 1997 & HIRES & 3600  & \raisebox{1.5ex}[0pt]{400} \\
 & & 11 August 1998 & Hamilton & 2700 & \\
 & & 12 August 1998 & Hamilton & 1800 & \raisebox{1.5ex}[0pt]{125}\\
HD 216143 & \phn 7.82 &  11 August 1998 & Hamilton & 1800 & 125 \\
HD 218857 & \phn 8.95 &  11 August 1998 & Hamilton & 2700 & \phn 75 \\
BD -18 5550 & \phn 9.29 & 12 May 1997 & HIRES & 1800 & \\
 & & 20 June 1997 & HIRES & 1800 & \raisebox{1.5ex}[0pt]{350} \\
 & & 12 August 1998 & Hamilton & 2700 & \phn 90 \\
BD -17 6036 & 10.52 & 12 May 1997 & HIRES & 1200 & 130 \\
 & & 11 August 1998 & Hamilton & 3600 & \\
 & & 12 August 1998 & Hamilton & 3600 & \raisebox{1.5ex}[0pt]{\phn 75} \\
BD -11 145 & 10.81 &  11 August 1998 & Hamilton & 4000 & \phn 60 \\
BD +4 2621  & \phn 9.98 & 12 May 1997 & HIRES & \phn 611 & \\
 & & 20 June 1997 & HIRES & 2400 & \raisebox{1.5ex}[0pt]{350} \\
BD +5 3098  & 10.55 & 20 June 1997 & HIRES & \phn 600 & 120 \\
 & & 11 August 1998 & Hamilton & 2700 & \\
 & & 12 August 1998 & Hamilton & 2700 & \raisebox{1.5ex}[0pt]{100} \\
BD +8 2856  & 10.07 & 20 June 1997 & HIRES & 2700 & 200 \\
 & & 8 March 1999 & Hamilton & 3600 & \phn 45 \\
BD +9 3223 & \phn 9.27 & 12 August 1998 & Hamilton & 2000 & 100 \\
BD +10 2495 & \phn 9.72 & 7 March 1999 & Hamilton & 3600 & \phn 75 \\ 
BD +17 3248 & \phn 9.40 & 12 August 1998 & Hamilton & 1800 & \phn 80 \\
BD +18 2890 & \phn 9.84 & 8 March 1999 & Hamilton & 3600 & \phn 70 \\
M92 VII-18 & 12.18\tablenotemark{1} & 12 May 1997 & HIRES & 3600 & \\
 & & 20 June 1997 & HIRES & 1800 & \raisebox{1.5ex}[0pt]{150} \\
\enddata
\tablenotetext{1}{$V_{\rm mag}$ from Shetrone 1996}
\end{deluxetable}

\subsection{Data Reduction}

The data were reduced using standard IRAF\footnote{IRAF is distributed
by the National Optical Astronomy Observatories, which are operated by
the Association of Universities for Research in Astronomy, Inc., under
cooperative agreement with the National Science Foundation} packages.
The data were corrected using the overscan region, and bias and flatfield
calibration frames.  The
Hamilton object spectra also required the subtraction of scattered
light.  Next, the spectra were extracted with variance-weighting and
3-$\sigma$ clipping, which also eliminated most of the cosmic rays.
The wavelength solution was derived from Th-Ar spectra.  For HIRES, a
rms of 0.002 \AA{} was achieved and for the Hamilton, a rms of 0.005
\AA.  If two or more spectra were obtained for the same object with
the same instrument, they were averaged.  For stars that had spectra
taken in both May and June, it was necessary to correct for the
different Doppler shifts due to the orbit of the Earth.  The radial
velocity shift was found using cross-correlation and the June spectra
corrected to the May data reference frame.

\subsection{Equivalent Widths}
We used the program SPECTRE (Sneden, private communication) to fit the
continuum and to measure equivalent widths (EWs) of unblended absorption
lines of interest.  The continuum fitting was done interactively by
marking continuum regions on a order which were then fit by a cubic
spline.  The EWs were determined by Gaussian fitting in most 
cases, although occasionally, such as lines with large wings, Simpson's
rule integration was used to measure the EW.  We checked the accuracy of 
EWs in three ways.  First, the echelle orders overlap in
wavelength for both HIRES and Hamilton data.  This overlap means
we have two independent measures of the EW of some of our lines.
In Figure 2, we show the difference in EW for the
same line measured on different orders.  The rms scatter is 1.6 m\AA{}
for 387 pairs of lines for the HIRES data and 2.9 m\AA{} for 3189
pairs of lines for the Hamilton data. 
For another internal comparison, we analyzed separately
 the two spectra of BD $-$18 5550 that were averaged for the final analysis.
These had the same exposure time (1800s) and so similar S/N.  We find
an averaged offset of 0.5 m\AA{} between the two sets of 232 EWs we measured.  
However,
that number is dominated by a few large differences at large EWs; if we
restrict the comparison to stars with EW $<$ 50 m\AA, the average difference is
$<$0.005 m\AA.  The rms variation regardless of EW limits is 
2.5 m\AA.   We also compared the EWs measured from the Hamilton data with those
from the HIRES data for stars that had been observed
with both.  We find $\langle EW_{Hamilton}-EW_{HIRES} \rangle =
-0.5 \pm 0.1$ m\AA.  This comparison involved 480 pairs of 
lines between 4200 \AA{} and
4700 \AA{}, and there were no trends with wavelength or EW.  
In our subsequent analysis, we created a 
master list for each star which included all the HIRES
EWs as well as Hamilton EWs for lines that were not covered in the HIRES
spectra.
We never averaged
EWs from the two spectrographs together, since the S/N of the HIRES data in the region of overlap was
always much higher than that of the Hamilton data.  
Table 2 gives our EWs.  EWs with wavelength shorter than 4710 \AA{} were
measured from HIRES data, while the rest come from Hamilton data.

\begin{figure}[htb]
\begin{center}
\includegraphics[width=3.0in,angle=0]{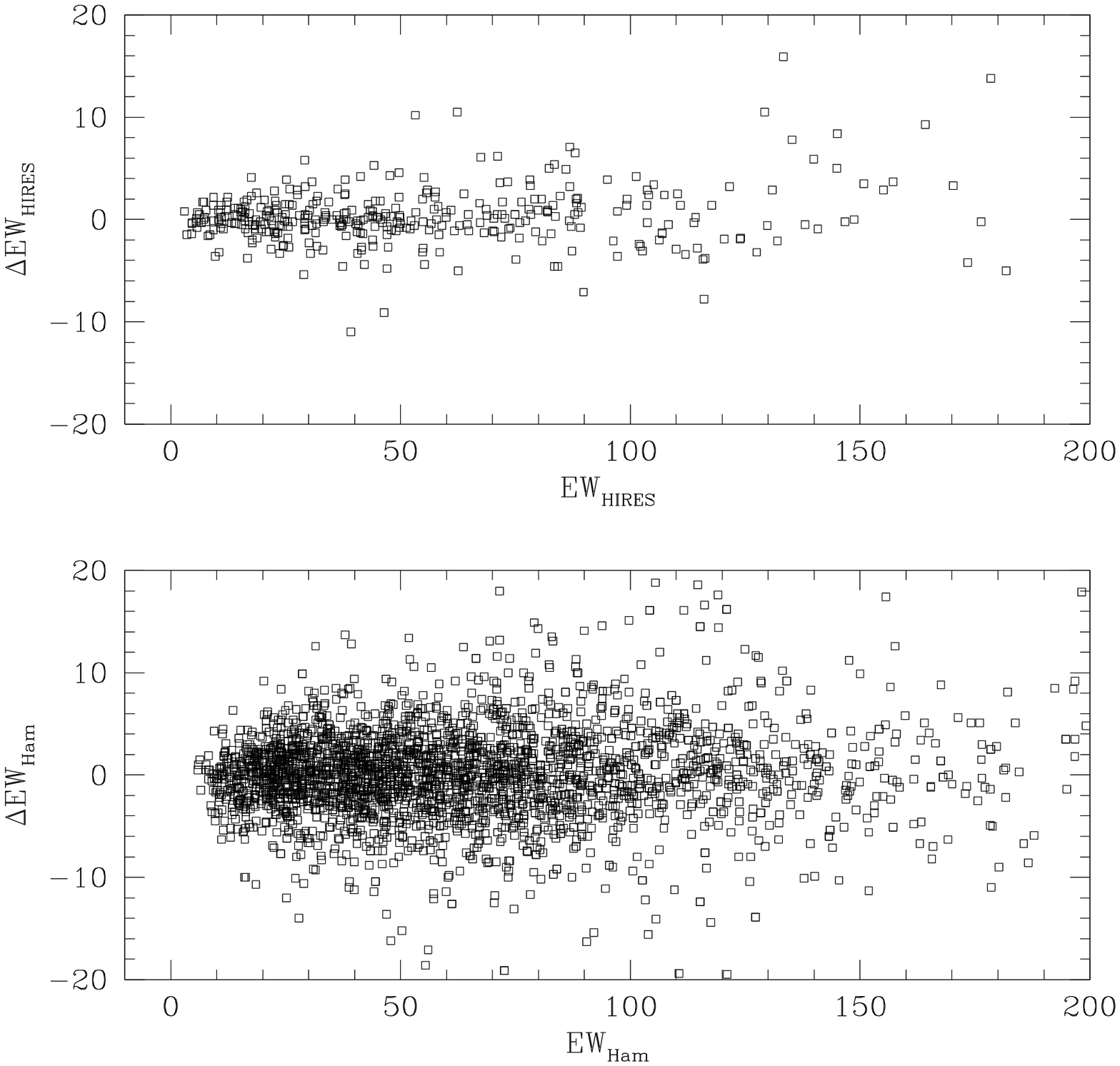}
\caption{$\Delta$ EW for lines re-measured in adjacent orders for 
(top) HIRES and (bottom) Hamilton}
\end{center}
\end{figure}

\subsection{Comparison with Previous Measurements}

Our list contains some well-known, well-studied metal-poor giants.  In
Figures 3a and b, we compare 
our EW measurements to some of the
recent studies of these stars.  The stars we have in common with each
study and the average offsets, standard errors of the mean 
and r.m.s. are listed in Table 3.

\begin{deluxetable}{llrrc}
\tablecaption{Stars for EW Comparison}
\tablewidth{0pt}
\tablenum{3}
\tablehead{\colhead{Reference} & \colhead{Stars in} & \colhead
{$\langle EW_{previous}-$} & R.M.S. & Number of  \\
\colhead{} & \colhead{Common} & \colhead{$EW_{this study}\rangle$} 
& & \colhead{EW pairs}
}
\startdata
Sneden \& Parsarthasay 1983 & HD 122563 & $-4.5 \pm 3.9$ m\AA & 21.1 m\AA &29\\
Gilroy \etal{} 1988 & HD 88609 & $+2.6 \pm 0.8$ m\AA & 3.9 m\AA & 23\\
 & HD 115444 & & &\\
 & HD 122563 & & &\\
 & HD 128279 & & &\\
 & HD 165195 & & &\\
Gratton \& Sneden 1988 & HD 216143 & $+10.8\pm 0.9$ m\AA & 9.5 m\AA & 123 \\
 & BD -18 5550 & & & \\
 & BD -17 6036 & & &\\
Gratton \& Sneden 1990,94 & HD 122563 & $+1.3 \pm 0.4$ m\AA & 2.1 m\AA &35 \\
 & HD 126587 & & &\\
 & HD 165195 & & &\\
Peterson \& Carney 1989 & HD 122563 & $+11.8 \pm 0.6$ m\AA & 10.0 m\AA & 110\\
\hspace{.2in}photographic & HD 128279 & &\\
 & BD -18 5550 & & &\\
\hspace{.2in}CCD & HD 122563 & $-1.2 \pm 0.9$ m\AA & 2.7 m\AA & 10\\
McWilliam \etal{} (1995a) & HD 126587 & $+2.8 \pm 0.3$ m\AA & 6.8 m\AA & 562 \\
 & HD 128279 & & &\\
 & HD 186478 & & &\\
 & BD -18 5550 & & &\\
Westin \etal{} 2000 & HD 115444 & $+2.0 \pm 0.1$ m\AA & 1.9 m\AA & 287\\
 & HD 122563 & & &\\
\enddata
\end{deluxetable}

Sneden \& Parsarthasay (1983) used silicon diode arrays to obtain
extensive data on the bright metal-poor giant HD 122563.  Their
S/N ranged from $\sim$ 100 in the blue to greater than 200 in the red.
Gilroy \etal{} (1988) measured EWs for a range of elements in giants, including
many of the neutron-capture elements we are interested in here.  Their
data had S/N$\sim$100 and R$\sim$30,000.  We agree well with both studies.
 Gratton \& Sneden (1988, 1990, 1994)
have published an extensive set of EWs for metal-poor stars.  
Gratton \& Sneden (1988) list EWs for
elements from Na to Ba derived from CCD spectra with R$\sim$20,000 and S/N
ranging from 80 to 200.  They also measured EWs for the light 
and
iron-group elements from CCD echelle spectra with higher resolution
(R$\sim$50,000) and S/N (S/N$>$ 150) (Gratton \& Sneden 1990). 
They have studied the neutron-capture elements with spectra of similar
resolution and S/N (Gratton \& Sneden 1994).  As seen in
Figure 3a, our EWs agree very well with the higher
resolution, higher S/N data,
but disagree with Gratton \& Sneden
(1988). Our disagreement increases with increasing EW.  We
find a similar disagreement with Peterson \& Carney
(1989) when considering their EWs measured from photographic plates,
and a similar improvement when comparing data for HD 122563 when
both sets come from observations with CCDs.  
\begin{figure}[htb]
\begin{center}
\includegraphics[width=3.0in,angle=0]{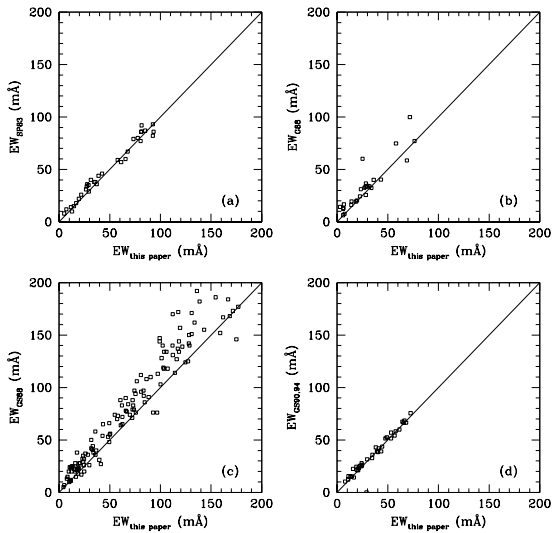}
\caption{EW comparison between our data and
 (a) Sneden \& Parthasarathy 1983
(b) Gilroy \etal{} 1988 (c) Gratton \& Sneden \etal{} 1988 (d) Gratton
\& Sneden 1990, 1994}
\end{center}
\end{figure}

\setcounter{figure}{2}

\begin{figure}[htb]
\begin{center}
\includegraphics[width=3.0in,angle=0]{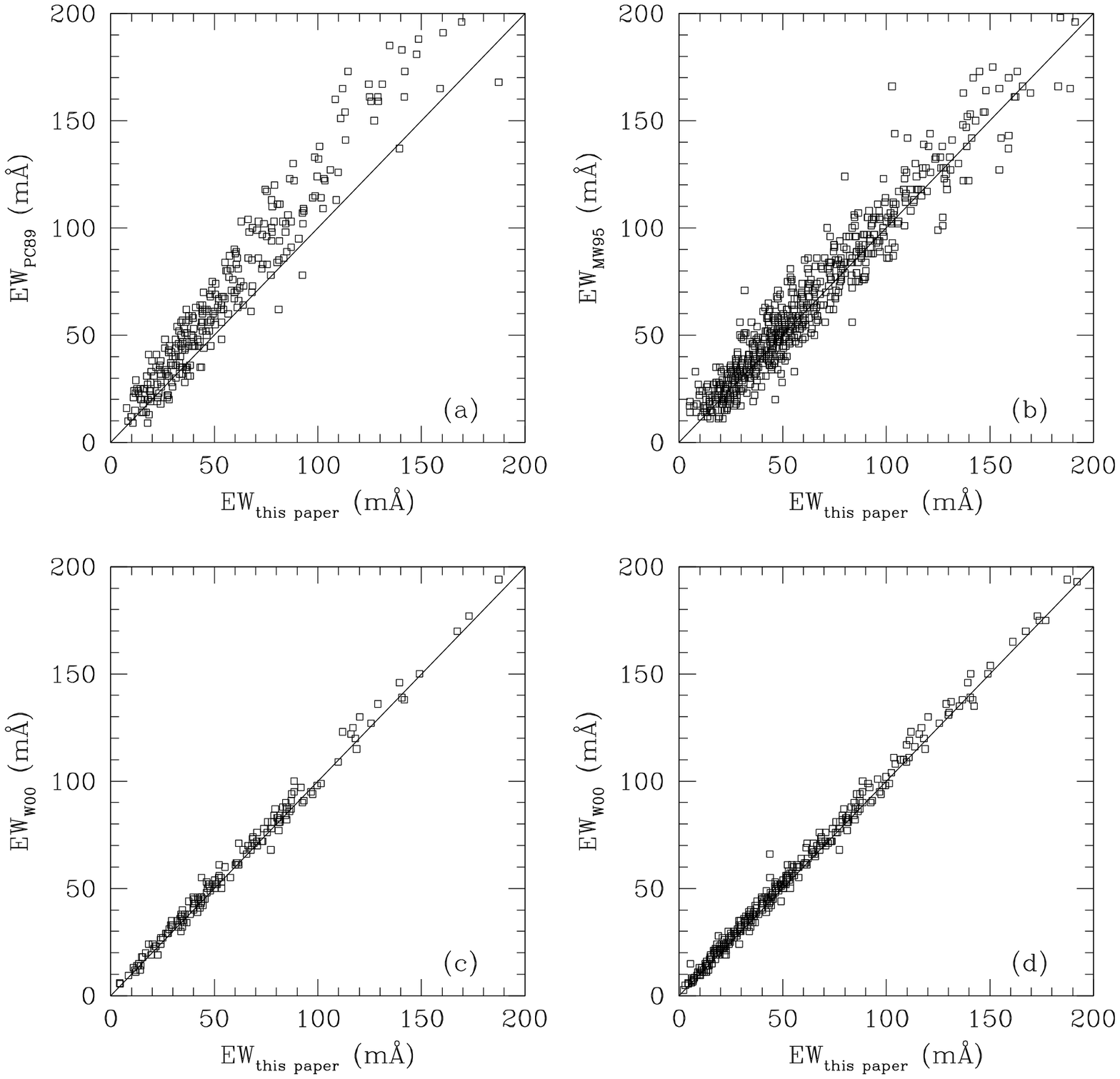}
\caption{EW comparison between our data and
 (a) Peterson \& Carney 1989 (photographic only) 
(b) McWilliam \etal{} 1995a (c) Westin \etal{} 2000 for HD 122563 (d)
Westin \etal{} 2000 for HD115444}
\end{center}
\end{figure}

McWilliam \etal{} (1995a) obtained spectra of very metal-poor stars with
the 2D-Frutti photon counting image device at Las Campanas.  Typical
S/N was 40 with R $\sim$ 22,000.  Their wavelength coverage was large,
extending from 3600 \AA{} in the blue to 7600 \AA{} in the red.  The 
average difference
in EWs is not large, 
but the low S/N of the McWilliam \etal{} (1995a) data leads
to the large scatter shown in Figure 3b.
The most interesting comparison is between our data and the recent work
of Westin \etal{} (2000).  They have very high S/N, very large wavelength
coverage, high resolution data for two bright metal-poor stars, HD 122563
and HD 115444.  They have measured EWs for a wide range of elements as well.
Figure 3b shows the comparisons for each of the stars.  
Our EWs are smaller by 2.0 m\AA, but the lack of scatter
is indicative of the high S/N and resolution of both data sets.

\section{Model Atmospheres}
\subsection{gf values}
Our \gf{} values were compiled from a variety of literature
sources (Table 2).  We used laboratory determinations whenever possible.  
However, \ion{Mg}{1} and
\ion{Ce}{2} lacked accurate laboratory values for many measurable lines, 
so we adopted values based
on theoretical calculations or analysis of the solar spectrum for these
two species.  If
there was more than one determination of an oscillator strength, we
attempted to determine which was the more accurate value. 
We consulted the authors' assessment of their errors as well as the critical
compilations of oscillator strengths 
of Fuhr \etal{} (1988) and Martin \etal{} (1988).
In a few cases, we averaged \gf{} values of approximately equal accuracy.
Our final selection of lines and oscillator strengths is summarized in
Table 2.  Those with more than one source are the average of the \gf{}
values found in those sources.  Below we list special notes on Fe, Ti and Cr.

{\ion{Fe}{1}:}  The most accurate relative oscillator strengths come from
the Oxford group (Blackwell \etal{} 1979a, 1979b, 1980, 1982a, 1982b,
 1982c).  However, they
normalized their data using one absolute \gf{} value (3719.94 \AA).  We 
compared their log \gf values to the log \gf{} values of Bard, Kock \&
Kock (1991) and O'Brian \etal{} (1991), which are of poorer relative
accuracy, but have been normalized using more lifetime measurements.
As found by other authors (e.g. McWilliam \etal{} 1995b), we found the
Oxford values to be systematically lower by 0.04 dex.  We have accordingly 
added 0.04 to the Oxford values. 

{Ti:}  The normalizations of the \ion{Ti}{1} and \ion{Ti}{2} \gf{} values have also been the
subject of some debate.  Grevesse \etal{} (1989) recommended that the
Oxford group's very accurate relative \ion{Ti}{1} values be increased by
0.056 dex because of new lifetime measurements by Rudolph \& Helbig (1982).
However, even these more accurate values gave a
solar Ti abundance (log$\epsilon$=4.99 $\pm$ 0.02) that disagreed with the
meteoritic Ti value (log$\epsilon$=4.93 
$\pm$ 0.02).  Bizzarri \etal{} (1993) measured
\ion{Ti}{2} transition probabilities by combining radiative lifetime
measurements with branching fractions.  They found that their log \gf{} values
were 0.093 higher than \ion{Ti}{2} log \gf{} values from the Oxford group,
which was not surprising given the uncertainty in the absolute scale of 
the Oxford log \gf{} values.  There was very
little scatter except for this offset.  Their comparisons with
Danzmann \& Kock (1980) and the theoretical calculations of Kurucz
(1988) showed considerably more scatter.  Our abundance analysis also
showed that the Bizarri \etal{} (1993) values resulted in smaller scatter in
the derived \ion{Ti}{2} abundance than
using Danzmann \& Kock (1980) or other values from the critical compilation
of Fuhr \etal{} (1988).  (The Oxford \gf{} values are for lines too far in
the blue to be of use).  Unfortunately, Bizzarri \etal{} (1993) found that their
\gf{} values resulted in a solar Ti abundance of log$\epsilon=5.04\pm0.04$, also in
disagreement with the meteoritic value, but in agreement with the \ion{Ti}{1} value
from solar analysis mentioned above.
Here we use the re-normalized \ion{Ti}{1} Oxford values, as suggested by Grevesse \etal
{} (1989)
and the Bizzarri \etal{} (1993) \ion{Ti}{2} values.  The uncertainty in our absolute
abundances  of 
Ti is $\sim$ 0.1 dex.

{\bf \ion{Cr}{2}:} The experimental results for \ion{Cr}{2} oscillator strengths
have been unsatisfactory.  The carbon arc measurements of Wujec
\& Weniger (1981) provided \gf{} values for lines in the optical part of
the spectrum.  However their normalization was very uncertain, since
the line they used was strongly affected by calibration problems.
Martin \etal{} (1988) recommended adjusting Wujec \& Weniger (1981)'s
 \gf{} values down by
$-0.84$ dex, based on a comparison with the theoretical values of
Kurucz and Peytremann (1975).  Recently, Pinnington \etal{} (1997a)
determined accurate lifetimes for some \ion{Cr}{2} levels based on selective
laser excitation.  However, in order to translate the lifetimes into
\gf{} values, they had to use the theoretical branching ratios of Kurucz (1988).
They used these \gf{} values to determine the solar Cr abundance and
found log$\epsilon=5.74\pm 0.06$, 
in agreement with the meteoritic value of
log$\epsilon=5.68\pm 0.03$.  Comparing the Pinnington 
\etal{} \gf{} values to the renormalized
Wujec \& Weniger \gf{} values, we find no systematic offset.  Therefore, we
used the renormalized Wujec \& Weniger \gf{} values for our two \ion{Cr}{2} lines.

\begin{deluxetable}{lcccccc} 
\tablenum{4}
\tablewidth{0pt}
\tablecaption{Colors and Reddenings}
\tablehead{\colhead{Star} & \colhead{$(V-R)$} & \colhead{$(V-K)$} 
& \colhead{E$_{(b-y)}$} &\colhead{E$_{(B-V)}$}  & \colhead{$(V-R)_0$}
& \colhead{$(V-K)_0$} }
\startdata  
HD 29574 & 1.159 & \nodata & 0.036 & 0.05 & 1.121 & \nodata \\
HD 63791 & 0.798 & \nodata & \nodata &  0.05 & 0.759 & \nodata \\
HD 88609 & 0.852 & 2.528 & \nodata &  0.04 & 0.821 & 2.418 \\
HD 108577 & 0.670 & 2.068 & 0.015 &  0.00 & 0.654 & 2.068 \\
HD 115444 & 0.746 & 2.327 & \nodata & 0.00 & 0.746 & 2.327 \\
HD 122563 & 0.805 & 2.485 & \nodata & 0.00 & 0.805 & 2.485 \\
HD 126587 & \nodata & 2.386 & 0.058 & 0.05 & \nodata & 2.168 \\
HD 128279 & \nodata & 1.940 & 0.039 & 0.040 & \nodata & 1.793 \\
HD 165195 & 1.076 & 3.275 & 0.099 & 0.25 & 0.971 & 2.903 \\
HD 186478 & 0.893 & \nodata & 0.057 & 0.09 & 0.833 & \nodata \\
HD 216413 & 0.874 & \nodata & 0.013 & 0.04 & 0.860 & \nodata \\
HD 218857 & 0.687 & \nodata & 0.021 & 0.03 & 0.665 & \nodata \\
BD -18 5550 & 0.885 & 2.706 & 0.086 & 0.08 & 0.794 & 2.383 \\
BD -17 6036 & 0.760 & 2.333 & 0.036 & 0.06 & 0.722 & 2.197 \\
BD -11 145 & 0.771 & \nodata & \nodata & 0.03 & 0.748 & \nodata \\
BD +4 2621 & 0.764 & \nodata & 0.003 & 0.00 & 0.761 & \nodata \\
BD +5 3098 & 0.708 & \nodata & 0.028 & 0.04 & 0.678 & \nodata \\
BD +8 2856 & 0.834 & 2.654 & \nodata & 0.00 & 0.834 & 2.654 \\
BD +9 3223 & 0.625 & 1.936 & 0.041 & 0.05 &  0.582 & 1.782 \\
BD +10 2495 & 0.678 & 2.105 & 0.002 & 0.00 & 0.676 & 2.097 \\
BD +17 3248 & 0.638 & 2.006 & 0.040 & 0.06 & 0.596 & 1.856 \\
BD +18 2890 &  0.656 & 2.075 & \nodata & 0.00 & 0.656 & 2.075 \\
\enddata
\end{deluxetable}

\subsection{Model Atmosphere Parameters}

We interpolated our model atmospheres from the updated grid of Kurucz 
(2001)\footnote{http://cfaku5.harvard.edu/}.  
Initial estimates for \teff{} for each star were obtained
from $V-R$ and $V-K$ photometry and the calibrations of Stone (1983) and
Cohen, Frogel, \& Persson (1978) respectively.  
The $V-R$ photometry was taken directly from Stone.
$V$ and $K$ magnitudes are from Alonso, Arribas, \& Mart\'inez-Roger (1998) 
if possible, or from Bond (1980).  When available, we
adopted the reddening estimates of Anthony-Twarog \& Twarog (1994), which
are based on Stromgren photometry.  Otherwise we adopted the reddening
values derived by Bond (1980).  We adopted the conversion value between
E$_{b-y}$ and E$_{B-V}$ of 0.73 from Anthony-Twarog \& Twarog as well.
Conversion between E$_{B-V}$ and A$_R$, A$_V$ and A$_K$ were done using
coefficients from Cardelli \etal{} (1989).  The photometry is
summarized in Table 4.  We interpolated the
[Fe/H]=$-2.26$ fiducial of Bergbusch \& VandenBerg (1992) 
at the appropriate \teff{}
to find our initial guess for log {\it g}.  [Fe/H] was taken from the estimate
of Bond (1980).
Next, we refined our initial estimates.  We used MOOG (Sneden 1973)
to determine LTE abundances.  We set the microturbent 
velocity ($\xi$) by requiring there be no dependence of the 
derived abundance from a line on its reduced
EW (RW=EW/$\lambda$) for \ion{Fe}{1}, \ion{Ca}{1}, \ion{Cr}{1} and \ion{Ti}{2}.   While many of the
elements showed no trend in abundance as a function of logRW at our adopted
$\xi$, some elements showed trends that changes of
$\sim \pm 0.3$ km/s in $\xi$ eliminated.  We have chosen $\pm 0.3$ km/s as our error
in $\xi$.  Magain (1984) argued that $\xi$ will be overestimated if
this method is used.  Some EWs will randomly be measured high, and this
will lead to higher abundances, thereby introducing a slope solely due
to random errors.  
He proposed using the {\it expected}
EW as the x-axis, which is derived using one abundance for all lines of 
a certain element,
and thus eliminates the correlation.  
Magain's method introduces its own bias with a
slope in the opposite sense
because both the expected EW and the abundance
derived from the observed EW depend on the \gf{} value of the line used.
A high \gf{} value will produce a large abundance, but the expected EW will
be systematically smaller than the observed EW, since a smaller average
abundance has been adopted.  From the curve of growth, we can see
that the bias of Magain's solution will be as large as the bias he is
eliminating when $\delta$ log \gf{} $\approx$ $\delta$ logRW.  So an
error of 0.05 dex in log \gf will have the same impact as 10\% error in
EW.  Errors of that magnitude are quoted for the O'Brian \etal{} (1991)
and Bard \etal{} (1991) \ion{Fe}{1} \gf{} values.  
The Oxford group's \gf{} values have smaller quoted
errors, but in \S 3.1, we renormalized them by 0.04 dex.  Our error
analysis in \S 2.3 and \S 2.4 shows that the errors in our EWs are $<$ 2 m\AA{} on average.
So we are in the regime where, depending on the strength of the line
and the accuracy of that particular \gf{} value, using the expected EW
could cause more of a bias than using the observed EW.  Since both
errors in EW and in log \gf{} value are \ltsim 10\%, any bias introduced
should be small.
The empirical bottom line is that we tested Magain's method on our
stars, and found that it did not affect our choice of $\xi$.

    \teff{} was changed until there
was no trend in the abundance versus excitation potential (E.P.) plot of the
\ion{Fe}{1} lines.  We estimate, based on the range of \teff{} that produce acceptable 
fits, that our errors are $\pm 100$K in T$_{\rm eff}$.  Next, we 
determined \logg by matching the \ion{Fe}{1} and \ion{Fe}{2} abundances.  While this could
be done precisely, we note that we have only $\sim$ 15 \ion{Fe}{2} lines,
so our \ion{Fe}{2} abundances have with a standard error of the 
mean $\sim$ 0.05.  The \logg values are affected by our renormalization
of many of our \ion{Fe}{1} lines (see above) as well.  Also, our gravities depend on our choice of temperature and
$\xi$.  Taking these effects into account, 
we found an acceptable range in \logg of $\pm$ 0.3 dex when attempting to find
a consistent model atmosphere. 
We note that because of the lower S/N data for M92 as well as
the lack of data in the red, our model atmosphere parameters are less
certain.  We have adopted $\pm$ 200 K, $\pm$ 0.4 dex and $\pm$ 0.3 km/s
as our errors for this star.  
Usually, two to three iterations on the model parameters were required
before the constraints on $\xi$, T$_{\rm eff}$, log {\it g}, and 
[Fe/H] were satisfied simultaneously.  We will refer to \teff, \logg,
and $\xi$ chosen by looking at Fe abundances as ``spectroscopic''.
Our choices for model atmosphere parameters are summarized in 
Table 5.

\begin{deluxetable}{ccccccccc} 
\tablenum{5}
\tablewidth{0pt}
\tablecaption{Model Atmosphere Parameters}
\tablehead{\colhead{Star} & \colhead{\teff} & \colhead
{\logg} & \colhead{[Fe/H]$_{mod}$} &\colhead{$\xi$}  & \colhead{\teff$_{phot}$}
& \colhead{\teff$_{IRFM}$} & \colhead{\logg$_{M92}$} & \colhead{\logg$_{M15}$}}
\startdata
HD 29574  &  4350 & 0.30 & $-$1.70 & 2.30 & 3950  & \nodata & \nodata & \nodata\nl 
HD 63791 &   4750 & 1.60 & $-$1.60 & 1.70 & 4725 & \nodata & \nodata & \nodata \nl
HD 88609 &   4400 & 0.40 & $-$2.80 & 2.40 & 4650 & 4600 & 0.94 & 0.83 \nl
HD 108577 & 4900 & 1.10 & $-$2.20 &  2.10 & 5050 & 5020 & \nodata & \nodata \nl 
HD 115444 &  4500 & 0.70 & $-$3.00 & 2.25 & 4775 & 4721 & 1.05 & 0.90\nl 
HD 122563 &  4450 & 0.50 & $-$2.65 & 2.30 & 4625 & 4572 & 0.96 & 0.85\nl 
HD 126587 &  4675 & 1.25 & $-$2.90 & 1.90 & 4950 & 4794 & \nodata & \nodata \nl 
HD 128279 &  5100 & 2.70 & $-$2.20 & 1.40 & 5325 & 5290 & \nodata & \nodata\nl
HD 165195 &  4375 & 0.30 & $-$2.20 & 2.50 & 4275 & 4237 & 0.76 & 0.60 \nl
HD 186478 & 4525 & 0.85 & $-$2.40 & 2.00 & 4550 & \nodata & 1.15 & 0.91 \nl
HD 216143 &  4500 & 0.70 & $-$2.10 & 2.10 & 4500 & \nodata & 1.05 & 0.90 \nl 
HD 218857 & 4850 & 1.80 & $-$2.00 & 1.50 & 4975  & \nodata &\nodata & \nodata \nl 
BD -11 145 &   4650 & 0.70 & $-$2.30 & 2.00 & 4750 & \nodata  & \nodata & \nodata \nl
BD -17 6036 &   4700 & 1.35&  $-$2.60 & 1.90 & 4850 & 4860 &\nodata & \nodata \nl 
BD -18 5550 &  4600 & 0.95 & $-$2.90 & 1.90 & 4700 & 4668 & 1.37 & 0.97 \nl 
BD +4 2621  &  4650 & 1.20 & $-$2.35 & 1.80 & 4725 & 5103 & \nodata & \nodata \nl 
BD +5 3098  &  4700 & 1.30 & $-$2.55 & 1.75 & 4925 & 4881  & \nodata & \nodata \nl 
BD +8 2856  &  4550 & 0.70 & $-$2.00 & 2.20 & 4525 & 4514 & 1.20 & 0.94 \nl
BD +9 3223 &  5250 & 1.65 & $-$2.10 & 2.00 & 5275 & 5363 & \nodata &\nodata \nl 
BD +10 2495 &  4900 & 1.90 & $-$2.00 & 1.60 & 4973 & 4939 & \nodata & \nodata \nl 
BD +17 3248 &  5200 & 1.80 & $-$1.95 & 1.90 & 5200 & 5236 & \nodata & \nodata \nl
BD +18 2890 & 4900 & 2.00 & $-$1.60 & 1.50 & 5000 & 5057 & \nodata & \nodata\nl 
M92 VII-18 & 4250 & 0.20 & $-$2.18 & 2.30 & &  & 0.64 & 0.45\nl  
\enddata
\end{deluxetable}

Our abundances are very insensitive to the [Fe/H] of the model atmosphere.  We
found that changing [Fe/H]$_{mod}$ by 0.2 dex changed the abundances by
$\sim 0.02$ dex.  However, the Kurucz models we used were made using scaled 
solar abundances.  In reality, most
metal-poor stars, including ours (\S 4) have enhanced ratios of the 
$\alpha$ elements.  Therefore
assigning a metallicity to the atmosphere based on [Fe/H] is incorrect.
To help account for this, we interpolated a model with
[Fe/H] $\sim$ 0.15 greater than the [Fe/H] determined from our lines.
The Kurucz models were created using log$\epsilon_{Fe}$=7.67 for the sun, a value
now considered to be at least 0.15 dex too large (see e.g. Bi\'emont \etal{} 1991; Asplund \etal{} 2000b).  
The combination of the
extra electrons from setting the overall metallicity too high and the extra 
electrons from Fe, should help account for the extra
electrons contributed by the $\alpha$ elements in metal-poor stars.
To compare with models with correct [$\alpha$/Fe], we 
used ATLAS9 to generate $\alpha$-enhanced models for three stars.
  We found changes of $\sim$ 0.01-0.02 (Table 6).  We
conclude that the $\alpha$-enhancement of the models is
not a critical source of error, given the uncertainties already
present in the models because of interpolation between grids of models
or grids
of opacities and choosing a Kurucz instead of MARCS model.

\begin{deluxetable}{lrrrrrrrrrrr}
\tablewidth{0pt}
\tablenum{6}
\tablecaption{log $\epsilon_{revised}-$log $\epsilon_{adopted}$  for systematic changes in Model Atmospheres}
\tablehead{
\colhead{Element} 
 & \colhead{$\Delta$ log $\epsilon$} 
 & \colhead{$\Delta$ log $\epsilon$}
 & \colhead{$\Delta$ log $\epsilon$} 
 & \colhead{$\Delta$ log $\epsilon$} 
 & \colhead{$\Delta$ log $\epsilon$}
 & \colhead{$\Delta$ log $\epsilon$} 
 & \colhead{$\Delta$ log $\epsilon$} 
 & \colhead{$\Delta$ log $\epsilon$}
 & \colhead{$\Delta$ log $\epsilon$}  
 & \colhead{$\Delta$ log $\epsilon$}  
 & \colhead{$\Delta$ log $\epsilon$} \\
\colhead{} & \colhead{$\alpha$}  &  \colhead{MARCS} & \colhead{NLTE}  
& \colhead{nover} 
& \colhead{$\alpha$}  &  \colhead{MARCS} & \colhead{NLTE} &\colhead{nover} 
& \colhead{$\alpha$}  &  \colhead{MARCS} & \colhead{NLTE}  
}
\startdata
 & \multicolumn{4}{c}{HD186478} & \multicolumn{4}{c}{HD128279} & 
\multicolumn{3}{c}{HD115444} \\
NaI & $-$0.04 & $-$0.12 & $-$0.10 &  $-$0.08 & 0.01 & 0.01 & $-$0.13 & $-$0.12 & 0.02 & $-$0.21 & $-$0.06 \\
MgI & $-$0.02 & $-$0.11 & $-$0.14 & $-$0.10 & 0.00 & 0.18 & $-$0.13 &$-$0.15& 0.03 & $-$0.15 & $-$0.21 \\
AlI & $-$0.05 & $-$0.07 & $-$0.18 & $-$0.10 & 0.01 & $-$0.06 & $-$0.11 & $-$0.17 & 0.08 &$-$0.52 & $-$0.25 \\
SiI & $-$0.01 & $-$0.10 & $-$0.05 & $-$0.08 & 0.00 & $-$0.04 & 0.00 & $-$0.12 & 0.00 & $-$0.12 & $-$0.12 \\
CaI &  $-$0.01 &  $-$0.07 &  $-$0.06 &  $-$0.08 & 0.00 &  $-$0.06 & $-$0.02 & 
$-$0.10 & $-$0.01 &  $-$0.08 &  $-$0.05 \\
ScII & $-$0.02 & $-$0.07 & 0.11 & $-$0.05 & $-$0.01 & 0.12 & 0.14 & $-$0.07 & 0.00 & 0.30 & 0.10 \\
TiI & $-$0.02 & $-$0.08 & $-$0.11 & $-$0.09 & 0.00 & $-$0.03 & $-$0.01 & 
$-$0.11 & 0.00 & $-$0.12 & $-$0.11 \\
TiII & $-$0.02 & $-$0.08 & 0.11 & $-$0.06 & $-$0.01 & 0.10 & 0.13 & $-$0.08 & 0.01 & 0.29 & 0.11 \\
VI & $-$0.01 & $-$0.07 & $-$0.11 & $-$0.09 & 0.00 & $-$0.02 & $-$0.01 & $-$0.11 & 0.00 & $-$0.11 & $-$0.10 \\
VII & $-$0.01 & $-$0.10 & 0.01 & $-$0.04 & $-$0.01 & 0.09 & 0.13 & $-$0.09 & 0.02 &  0.02 & $-$0.06 \\
CrI & $-$0.02 & $-$0.07 & $-$0.09 & $-$0.09 & 0.00 & $-$0.03 & $-$0.01 & $-$0.10 & $-$0.01 & $-$0.06 & $-$0.08 \\
CrII & $-$0.01 & $-$0.10 & 0.12 & $-$0.07 & $-$0.02 & $-$0.13 & 0.14 & $-$0.08 & $-$0.01 & $-$0.13 & 0.12 \\
MnI & $-$0.01 & $-$0.08 & $-$0.10 & $-$0.09 & 0.00 & $-$0.04 & $-$0.01 & $-$0.11 & $-$0.01 & $-$0.16 & $-$0.13 \\
MnII & $-$0.01 & $-$0.12 & $-$0.10 & 0.04 &  $-$0.01 &  0.11 & 0.11 & 
$-$0.10 & 0.06 & $-$0.17 & $-$0.17 \\
FeI & $-$0.02 & $-$0.07 & $-$0.09 & $-$0.09 & 0.00 & $-$0.03 & $-$0.03 & $-$0.12 &
0.01 & $-$0.08 & $-$0.09 \\
FeII & $-$0.01 & $-$0.09 & 0.12 & $-$0.05 & $-$0.02 & $-$0.02 & 0.14 & $-$0.08 
& $-$0.01 & 0.06 & 0.11 \\
CoI & $-$0.02 & $-$0.10 & $-$0.14 &$-$0.08 & 0.00 & $-$0.01 & 0.00 & $-$0.11& 0.00 & $-$0.17 & $-$0.14 \\
NiI & $-$0.02 & $-$0.04 &$-$ 0.09 & $-$0.8 & 0.00 & $-$0.04 & $-$0.03 & $-$0.13
& 0.08 & $-$0.27 &$-$0.21 \\
ZnI & $-$0.01 & $-$0.08 & 0.06 & $-$0.07 & \nodata & \nodata &  \nodata &  \nodata & $-$0.01 & 0.08 & 0.04 \\
YII & $-$0.02 & $-$0.07 & 0.06 & $-$0.03 & 0.00 & 0.16 & 0.14 & $-$0.07 & 
0.03 & 0.22 & 0.00 \\
ZrII & $-$0.01 & $-$0.09 & 0.03 & $-$0.02 & $-$0.01 & 0.15 & 0.14 & $-$0.07 &
0.01 & 0.15 & $-$0.02 \\
BaII & $-$0.03 & $-$0.05 & 0.16 & $-$0.04 & $-$0.01 & 0.20 & 0.13 & $-$0.07 & 0.03 & 0.59 & 0.15 \\
LaII & $-$0.01 & $-$0.08 & 0.11 & $-$0.05 & $-$0.01 & 0.15 & 0.14 & $-$0.07 & 
0.00 & 0.30 & 0.08 \\
CeII & $-$0.01 & $-$0.07 & 0.10 & $-$0.05 & \nodata & \nodata & \nodata & \nodata & 0.00 & 0.35 & 0.10 \\
PrII & \nodata & \nodata & \nodata & \nodata & \nodata & \nodata & \nodata &
\nodata & 0.00 & 0.29 & 0.07 \\
NdII & $-$0.01 & $-$0.07 & 0.11 & $-$0.05 & $-$0.01 & 0.15 & 0.14 & $-$0.07 & 
0.00 & 0.32 & 0.08 \\
SmII & $-$0.01 & $-$0.07 & 0.11 & $-$0.05 & \nodata & \nodata &  \nodata &  \nodata & 0.00 & 0.34 & 0.10 \\
EuII & $-$0.01 & $-$0.09 & 0.08 & $-$0.04 & $-$0.01 & 0.16 & 0.14 & $-$0.07 &
0.01 & 0.26 & 0.06 \\
GdII & $-$0.02 & $-$0.09 & 0.04 & $-$0.04 & \nodata & \nodata &  \nodata &  \nodata & 0.01 & 0.24 & 0.03 \\
TbII & \nodata & \nodata & \nodata & \nodata & \nodata & \nodata & \nodata
& \nodata & 0.00 & 0.26 & 0.06 \\
DyII & $-$0.01 & $-$0.08 & 0.04 & $-$0.03 & 0.00 & 0.19 & 0.14 & $-$0.07 & 
0.02 & 0.20 & $-$0.01 \\
ErII & $-$0.02 & $-$0.09 & 0.06 & $-$0.05 & $-$0.01 & 0.13 & 0.14 & $-$0.07 &
0.01 & 0.23 & 0.03 \\
TmII & \nodata & \nodata & \nodata & \nodata & \nodata & \nodata & \nodata &
\nodata & 0.01 & 0.22 & 0.05 \\
YbII & $-$0.02 & $-$0.09 & 0.02 & $-$0.03 &  $-$0.01 & 0.12 & 0.14 & $-$0.08 &
0.08 & 0.24 & $-$0.04 \\
\enddata
\end{deluxetable}

\subsection{Model Atmosphere Concerns}

\subsubsection{Spectroscopic vs. Photometric Temperatures}

The \teff{} values based on photometry were usually 100-150 K hotter than the
spectroscopic \teff{} values.  Figure 4 shows the plots 
of \ion{Fe}{1} abundances vs. E.P. of the lines for the
star HD 115444 for both its spectroscopic and photometric temperatures.
Clearly a slope is present when
the photometric temperature is adopted.
While the calibrations of Stone (1983) and Cohen \etal{} (1978)
give answers that 
differ by up to 100 K, they are consistently higher than the
spectroscopic temperatures.  The more recent calibrations by di Benedetto
 (1998) and 
Alonso, Arribas \& Mart\'inez-Roger (1999b) give similar answers.  The problem also cannot be 
attributed to errors in reddening since stars with assumed
$E(B-V)=0$ show this phenonmenon.  It is also not due to stars used for
calibration having different characteristics than our giants, since
all calibrations were based on metal-poor giants.  In fact
Alonso, Arribas, \& Mart\'inez-Roger (1999a) found temperatures for many of our stars based on the
Infrared Flux Method (IFRM) to determine their conversion from colors
to \teff.  We include those IFRM measurements in Table 5.
Finally, our spectroscopic temperatures could be wrong because of errors
in the \gf{} values that are correlated with E.P.  However, investigations
by Blackwell, Booth \& Petford (1984a) have shown that is not the case with the
Oxford values.  Restricting ourselves to just the Oxford values results
in the same spectroscopic \teff{}, a result we could anticipate by the
general good agreement between the Oxford and the other \ion{Fe}{1} 
\gf{} values we used.

\begin{figure}[htb]
\begin{center}
\includegraphics[width=3.0in,angle=0]{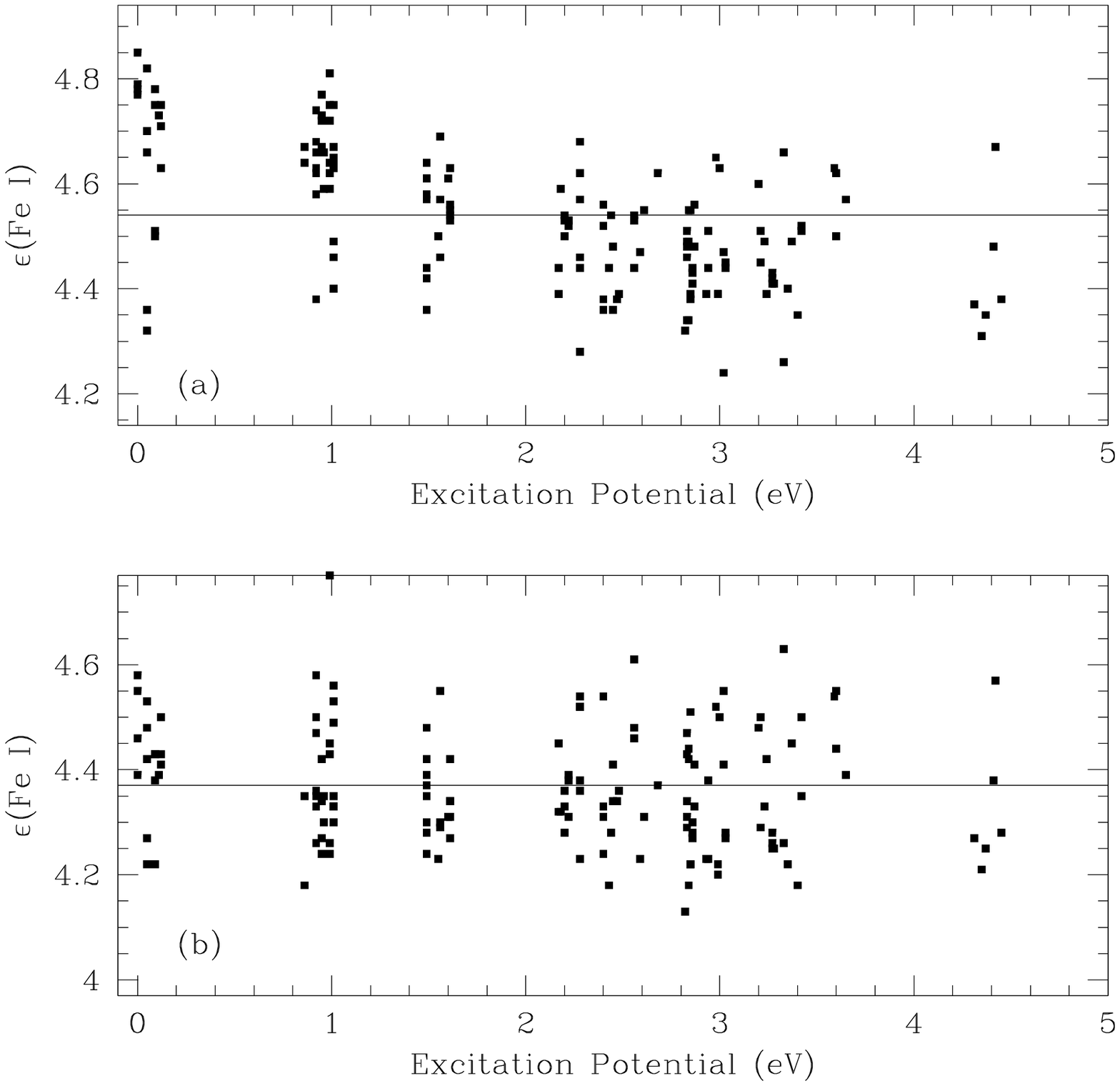}
\caption{Comparison of \ion{Fe}{1} abundance versus E.P. plots
for HD 115444 with (a) the photometric temperature of 4700K and 
(b) the spectroscopic temperature of 4500K}
\end{center}
\end{figure}
Although this systematic offset of $\sim$ 100K is a source of 
concern, it represents an error of only 2-3\% in our \teff.
The conversion of colors or infrared fluxes to \teff{} for metal-poor
stars is indirect and requires the use of model fluxes and bolometric
corrections, while \teff{} derived from E.P. plots requires
accurate temperatures throughout the model atmosphere and is most
sensitive to the flux in the visual wavelengths.  Given the uncertainties
in model atmospheres, it is not surprising that
the \teff{} derived with different methods only agree at the 3\% level.
First, we note that all color-\teff{} relations mentioned rely on atmosphere
models at some level.  Stone (1983) adopted temperatures from the literature
which were based on a variety of methods including \ion{Fe}{1} vs. E.P. plots 
(Luck \& Bond 1981) and a
theoretical $(V-K)-$\teff{} relation (Cohen \etal{} 1978).  This was the
$V-K$ relation we also used, and it is based on ATLAS6 model atmospheres
to establish the colors and bolometric corrections for a star with a given
\teff.  
The calibration of Alonso \etal{} (1999b) is based on IRFM which we discuss
next.

Because measurements of the bolometric luminosity of metal-poor giants
are nonexistent, stellar atmospheric models must be used to translate
a measured IR flux into a total flux, the key of the IRFM method.
Alonso \etal{} (1999a) saw that a variation of 5\% in the ratio of bolometric
flux to IR flux causes variations of 1.6\% in the \teff{} derived using
the K Band.   
M\'egessier (1994) found differences of 1.5\% in \teff{} at 
6000K comparing IRFM \teff{}
derived from Kurucz and MARCS models.  The treatment of convection within
the models also affects where the flux emerges (e.g. Castelli, Gratton,
\& Kurucz 1997).
Blackwell, Lynas-Gray \& Petford (1991) found that the IRFM \teff{} increased by 1\% when they
adopted a new zeropoint for their photometry and included improved H$-$ opacity
in MARCS models.  So the IFRM temperatures are not model-independent.
Clearly, neither are the spectroscopic \teff{} determinations.  
To check the effect of different atmosphere models on the dependence of \ion{Fe}{1} abundance and E.P., we created
MARCS (Bell \etal{} 1976) models for three stars, and determined their model atmosphere
parameters
using the same method as for the Kurucz models.  We found higher
spectroscopic temperatures, in much better agreement with the photometric temperatures,
but lower gravities as well, so the final models were well away from
theoretical isochrones.  Reassuringly, the derived abundances changed
by far smaller amounts than if we adopted our Kurucz model atmosphere 
parameters for
the MARCS models as well (Table 6).  In fact, except for HD 115444, the
relative abundances for the rare earths, for example, change by only $\sim$ 0.01-0.05 dex.  The large
scatter for HD115444 reflects more our inability to find a MARCS
model that simultaneously eliminated trends in the \ion{Fe}{1} abundance vs.
E.P. plot and gave identical \ion{Fe}{1} and \ion{Fe}{2} abundances.  The overall failure
of the spectroscopic and photometric temperatures to agree reflects the 
inability of models to simultaneously reproduce both the emerging flux and the 
shape and depth of absorption lines.  Which aspects of metal-poor stellar
atmospheres are closer to the truth is not
clear.  This is an instance where having model-independent radii, temperatures,
or distances for metal-poor giants would be invaluable.

We choose
as our ``\teff'' the model temperature which does not show a trend
in the derived Fe abundance vs. E.P. plot.  
This \teff{} may not give the right bolometric flux or color of the star,
but means that the derived abundances do not depend on the E.P. of
the lines measured.  This eliminates a potential bias in our relative
abundances as more metal-poor stars have fewer high E.P. lines 
with measurable EWs.

\subsubsection{Convection}
The treatment of convection in the atmosphere affects the temperatures
in the line-forming region, and that can change abundances by $\sim$ 0.1 dex
(e.g. Ryan \etal{} 1996).  
Castelli \etal{} (1997) explored the effect of convection
on the temperature structure and resulting flux, focusing in particular
on the use of ``approximate overshooting'' in Kurucz models.  
Castelli \etal{}  argued that the models without overshooting produced 
color-\teff{} relations that were in good agreement with stars with
known colors and \teff{} measured by IRFM (Blackwell \& Lynas-Gray 1994; Smalley
\& Dworetsky 1995).
On the other hand, the solar model with overshoot explained more observations
than its counterpart without overshooting.  Castelli (2001)\footnote{http://cfaku5.harvard.edu/} created
a set of models with the
``approximate overshooting'' of the ATLAS9 models turned off.  This
meant less energy deposition in the deepest layers, and therefore lower
temperatures.  These models are available for [Fe/H] $\geq -2.5$.  For
the two stars in Table 6 with [Fe/H] $> -2.5$, we have included the
changes in log $\epsilon$ when the models without overshooting are used.
We used the same model atmosphere parameters as for the original Kurucz
models, because these proved to be a good match with the data.
These are listed as $\Delta$ log $\epsilon$(NOVER).
As expected, it has larger effect on the absolute abundances than on
the relative abundances.

\subsubsection{Dependence of Abundance on Wavelength}

An examination of our abundance 
analysis of the most metal-poor stars in our sample, such as 
HD 115444, HD 88609, and HD 122563, revealed a correlation 
between derived abundance and wavelength of the line.  
For example, for HD 115444, 
the lines blueward of 4700 \AA{} produced an average deviation from the mean 
abundance of 0.03 dex, while the ones redward of 4700 \AA{} have an average
deviation of $-0.06$ dex.  The problem gets progressively worse the smaller the
wavelengths used.  If we had considered only lines with $\lambda < 4000$ \AA,
then the average deviation of the blue lines is 0.09 dex.  
The bluer lines also tend to produce larger abundances regardless of whether
only Hamilton or only HIRES data is used.  So this problem is not directly
attributable to combining HIRES and Hamilton data.  
One possible explanation is that the continuum was systematically overestimated
in the bluer regions, as the S/N decreased and the crowding
increased.  Continuum placement is certainly
contributing to the errors in our EWs, but it does not seem to be the
root of this discrepancy.  First, the more metal-rich stars
([Fe/H] $>-2.7$) show no difference
in abundance between the red and blue lines, although continuum placement 
should be even more problematic in the more crowded metal-rich spectra.
Second, we divided Westin \etal 's (2000) EWs into blue and
red regions and compared our EWs to them.  
We found for HD 115444 that our EWs from lines with 
$\lambda < 4700$ \AA{} were on average 4.35 m\AA{} smaller than the EWs of
Westin \etal, while the EWs from lines with $\lambda > 4700$ \AA{} were
much closer in magnitude, 
with ours only 0.25 m\AA{} smaller.  A comparison between the
two data sets for HD 122563 revealed much the same thing.  Our EWs were
systematically smaller, but the difference for the EWs 
from the blue lines was $\sim$ 1 m\AA{} larger than for the red lines. 
If we corrected our EWs for the offset between our and Westin \etal 's data,
the bluer lines would give even larger abundances than with our original
EWs.  However, we note that it takes appreciable errors in EWs to produce the deviations
seen.  Abundance differences of 0.09 dex would require errors of 25\% in the EWs for
lines in the linear curve of growth, and even larger errors for stronger lines.

A noticeable improvement in the agreement between blue and red lines
was achieved by using the hotter photometric temperature scale.  This
was not due to changes in \logg{} or a concentration of high or low
excitation lines in a certain wavelength range.  Using a model with
the photometric temperature, but the standard log {\it g}, eliminated
the blue-red discrepancy in the Fe I lines in HD 115444 even when only
lines with E.P.s between 2 and 3 eVs were considered.  Another method
for improving the problem was using MARCS models which, as noted
earlier, tended to have hotter \teff s, though coupled with very low
log {\it g}.  The ability of higher temperature models to improve the
situation provides additional evidence that the temperature structure
in metal-poor stars has not yet been accurately modeled (see \S 3.3.1
and 3.3.4 as well).  Another possible cause of the blue-red discrepancy is
too much continuous opacity in the models of very metal-poor stellar
atmospheres.  Short \& Lester (1994) found the opposite effect in Arcturus;
the Kurucz ATLAS 9 models produced lower abundances in the blue regions
than the red region.  They found that adding a additional continous
absorption opacity approximately equal to the opacity included in the
ATLAS 9 code solved the disagreement.  Perhaps the opposite effectis happening
in very metal-poor atmospheres.

This dependence on wavelength increases the error
in the abundances, but for many elements with lines spread throughout
our spectra, this error is less important that the errors in the
parameters of the model atmospheres.  For a few elements, such as Yb,
only lines in the blue were measured; therefore a small systematic
bias of $\sim$ 0.05 dex exists in the most metal-poor stars.  However,
the observational error for these elements usually outweighs this
effect; our assumed error for Yb is 0.20 dex, for example.
Examination of Table 2 shows which elements have only blue lines, and
Table 8 gives the r.m.s. scatter produced by the lines of an element,
so the susceptibility of each element to this effect can be judged.

\begin{figure}[htb]
\begin{center}
\includegraphics[width=3.0in,angle=0]{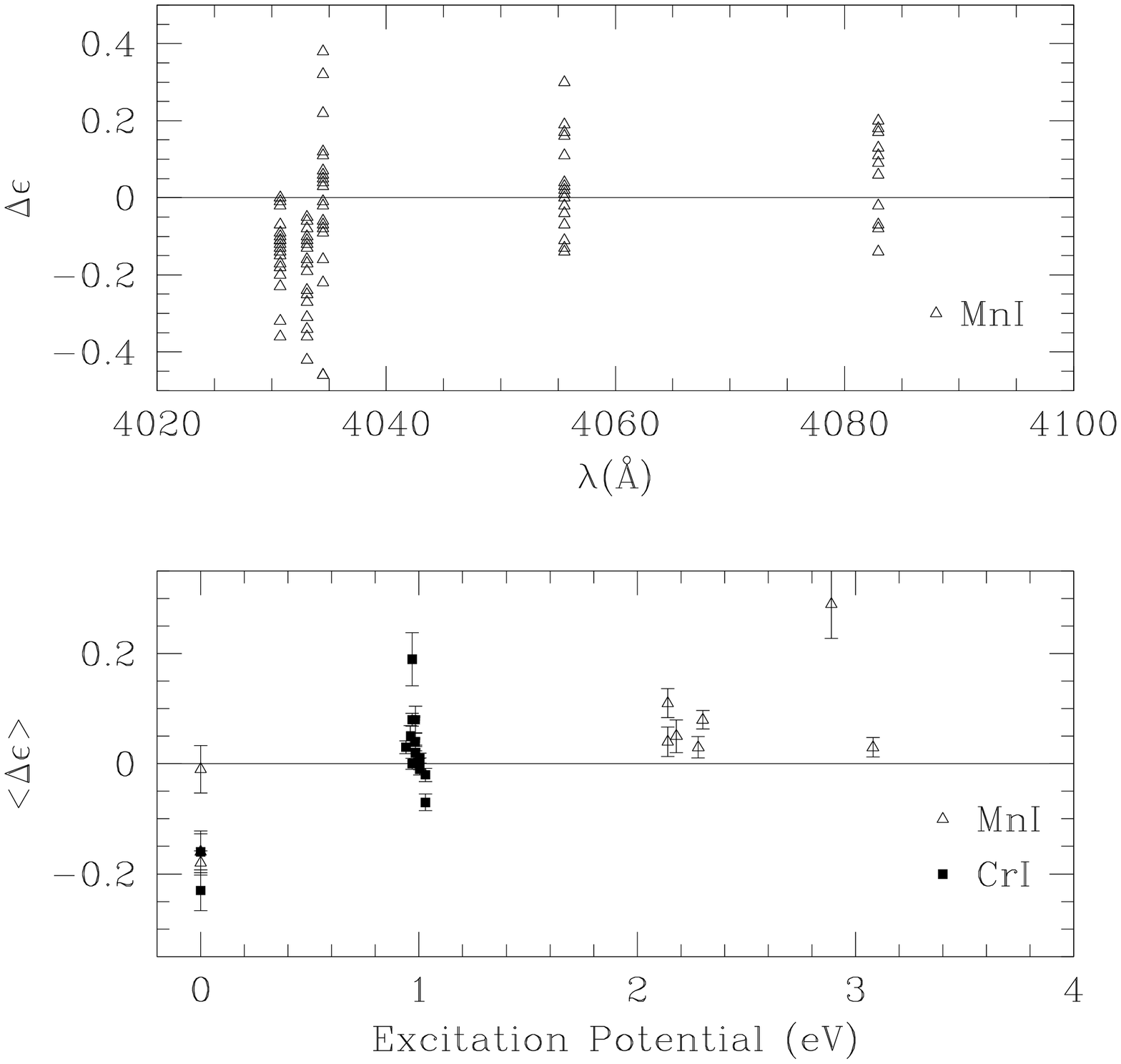}
\caption{Deviations from the mean abundance for \ion{Mn}{1} and \ion{Cr}{1} l
ines.  Top:
the individual data points for \ion{Mn}{1} lines.  Bottom:  the mean
deviations and errors for our \ion{Mn}{1} and \ion{Cr}{1} lines.}

\end{center}
\end{figure}

\subsubsection{Deviant Lines}

There are some
lines of \ion{Cr}{1}, \ion{Mn}{1} and \ion{Fe}{1} that systematically give lower abundances
in our analysis, but whose \gf{} values have been judged very good.  
The
deviant lines are the strongest lines of their species.
Figure 5 illustrates the problem for \ion{Cr}{1} and \ion{Mn}{1}.    
The resonance lines of these elements 
at 4000-4200 \AA{} give abundances that are lower
by $\sim$ 0.2-0.4 dex than weaker, higher excitation lines.  This appears
to be a separate problem from the correlation discussed in \S 3.3.3, 
where the bluer lines, regardless of EW, gave {\it higher} abundances in
the most metal-poor stars.  The strong lines always produce lower 
abundances, even for
the more metal-rich stars.  In fact, since the more metal-rich stars 
have stronger lines, more lines are affected as the metallicity increases.
If there were blending from additional, unidentified lines in the EWs, 
the
abundances would be too high.  NLTE corrections, 
at least for \ion{Fe}{1} would make the problem worse (Dalle Ore 1993;
Gratton \etal{} 1999).  
Adopting a
model without overshooting, an alpha-enhanced model, or a MARCS model
 does 
not solve the problem.  
Finally, if the lines are weaker than about 100-120 m\AA, as they are in the most
metal-poor stars, they do not give systematically low
abundances.  This suggests that
the problem lies in the upper layers of our model atmospheres, where
substantial parts of the affected lines are formed.  We considered
two possibilities:  depth-dependent $\xi$ and the temperature structure in
the outer atmosphere.
There is observational support for depth-dependent $\xi$ in
Arcturus (e.g. Gray 1981; Takeda 1992), but those authors found that
$\xi$ increased as $\tau$ decreased, the reverse of what is demanded
here.  However, we decided to take advantage of MOOG's ability to
handle depth-dependent $\xi$ to see if this was even a possible option.
We set $\xi$ for each layer in our HD 186478 atmosphere, beginning with
very low ($\xi$=1cm/s) in upper levels and gradually increasing
to $\xi$=2.0km/s, our best universal $\xi$ value by the middle layers.
We know that this is not completely
self consistent since the atmosphere was created using opacity functions 
that were not depth-dependent, but with stars this metal-poor, this
will not affect our judgement of the viability of this option.  
Figure 6 illustrates the result.  Deeper layers contribute heavily to 
the equivalent width for these very saturated lines, so $\xi$ needs
to be low in these layers as well to fully correct the low abundances.
However, somewhat weaker lines (80-100 m\AA{}) are mostly formed
in these layers as well, and are affected adversely by the low $\xi$.

\begin{figure}[htb]
\begin{center}
\includegraphics[width=3.0in,angle=0]{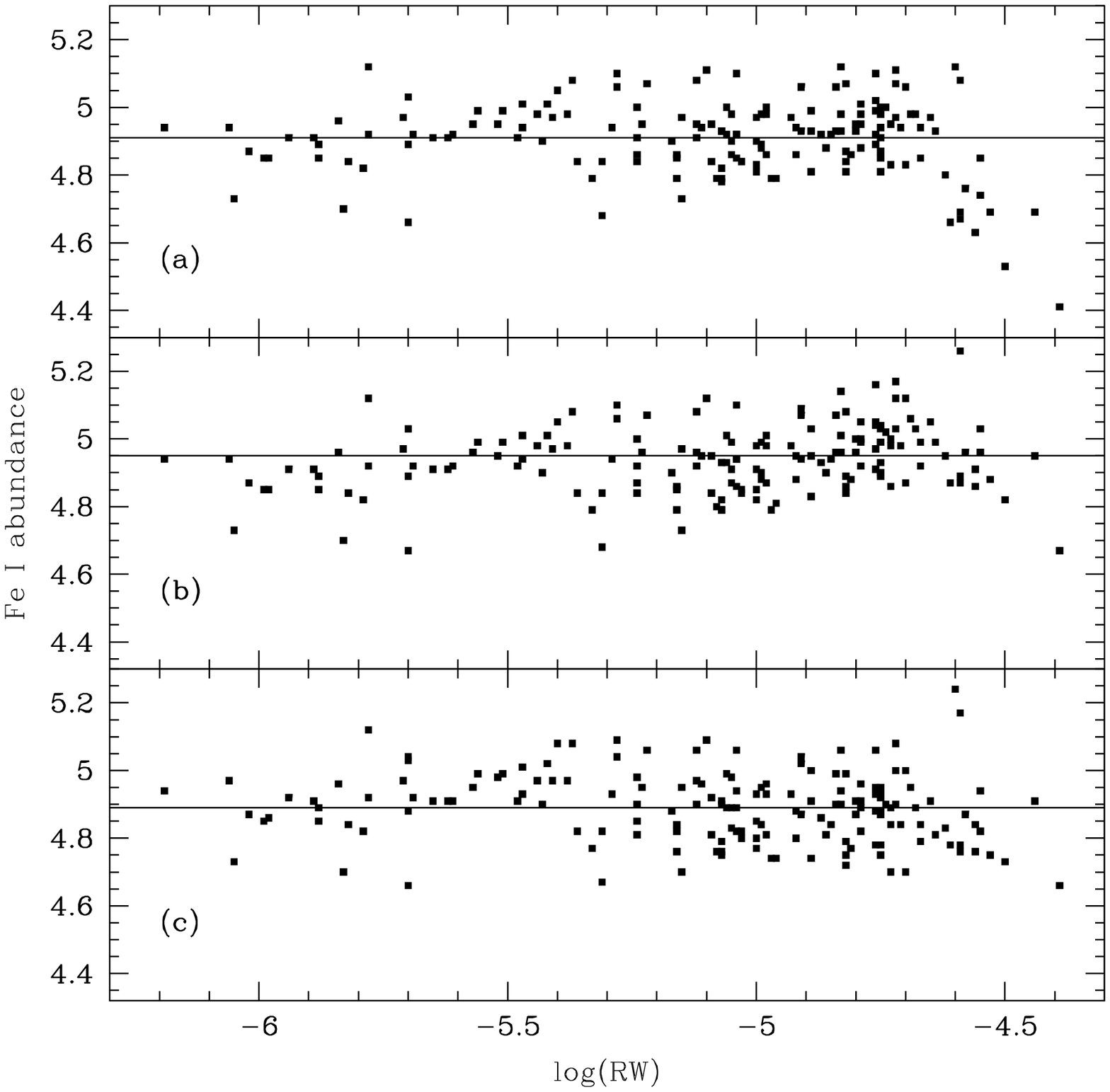}
\caption{\ion{Fe}{1} abundance vs. log(RW) for lines from HD186478 using (a)
the standard model (b) a model with depth-dependent $\xi$ and (c)
a model with T$_{min}$=0.75\teff.}  

\end{center}
\end{figure}
As for the other possibility, the upper layers of the Kurucz
model atmospheres are likely incorrect (McWilliam \etal{} 1995b).  
Dupree, Hartmann \& Smith (1990) showed that metal-poor giants have chromospheres. 
Empirical studies have shown that chromospheres have a 
minimum temperature (T$_{min}$) at 0.75\teff{} 
before rising steeply
to temperatures greater than 10,000 K (Kelch \etal{} 1980).  
The upper layers of Kurucz
models are instead characterized by low temperatures (T$<$0.50\teff).
These low temperature layers contribute much to the absorption of low
ionization, large EW lines.  To test the idea that the temperature structure
in the upper layers is at fault, we
created new models for HD 186478 by changing the T($\rho$x) relation in 
the upper layers of the original Kurucz atmosphere.  We adopted a linear
relation for the decrease of T, starting at the T($\rho$x) where deviant
lines alone had substantial contributions (log$\tau_{5000} \sim -1.2$) 
and ending at T=0.75\teff{} at the last layer.
Radiation pressure was assumed to be unimportant, and the gas
pressure determined from hydrostatic equilibrium.  The electron density
was calculated using a program kindly given to us by A. McWilliam.
using the formulism of Mihalas (1978).  Obviously this
model is unphysical, since flux is not conserved.  However,
the purpose of this exercise is to test the viability of such a 
solution and its effect on the abundances, rather than derive accurate
abundances.  The resulting model
produced trends in the \ion{Fe}{1} abundance vs. logRW plot.  After raising $\xi$ by 0.3
km/s to remove those trends, the
strong \ion{Fe}{1} lines give the same abundance as the rest of the lines (Figure 6).

The abundances of many elements change when this new model is used,
in part because of the increased $\xi$. 
However, while it represents a possible solution to the deviant lines,
this model cannot be used to calculate reliable abundances.  The true $\Delta$log$\epsilon$ is not certain, but our uncertainties in \teff{} and $\xi$ should 
reflect most of the effect.  Simply truncating the Kurucz models the first
time they dip below 0.75\teff{} still results in too low abundances for
the deviant lines.  This would make them similar to the MARCS models
which have fewer upper layers, but which also fail to improve the agreement
between weak and strong lines.  The upper layers are necessary for
this solution to work, but they must have higher temperatures.
For this analysis, we will continue to use the Kurucz models unmodified.
However, we have decided to exclude the resonance lines of \ion{Cr}{1} and \ion{Mn}{1}.  
For stars with lower metallicities or lower S/N data, 
fewer lines other than the resonance lines can be measured.
Therefore, if we include all lines, we would introduce an offset between
the more metal-poor end and more metal-rich end due to the increasing 
influence of the strong lines.  The
strong \ion{Fe}{1} lines could potentially affect our \teff{} determinations
from the \ion{Fe}{1} abundance versus E.P. plots.  We checked the
effect of removing the strong \ion{Fe}{1} lines from our EW list, and found that
while the scatter was noticeably reduced, the derived \teff{} did not change
($\sim$ 25 K at most).  This is because a new \teff{} value would affect 
the lines arising from 2 eV and higher levels as well, and those
are unaffected by the strong line problem.  Fe abundances overall are not
affected by these strong lines, since they are a small percentage of
the total lines, and they are included in the abundance determination.

\subsubsection{NLTE effects on \logg} There is, unfortunately, a systematic
error that may result from choosing our model atmosphere parameters
by ionization balance.  Two recent papers (Thevenin \& Idiart
1999; Allende Prieto \etal{} 1999) pointed out that the assumption of LTE
may be incorrect, and that consideration of non-LTE may change the
derived spectroscopic gravity.  \ion{Fe}{1} levels are depopulated by
ultraviolet radiation when non-LTE is considered, while \ion{Fe}{2} is
relatively unaffected.  Therefore an LTE analysis underestimates the
amount of \ion{Fe}{1} required in an atmosphere, resulting in \logg values
that are too low.  Allende Prieto \etal{} (1999) found much better agreement
with the Hipparcos gravities for subdwarfs when the NLTE gravities
were used.  Dalle Ore (1993) calculated the NLTE effects on \ion{Fe}{1} lines
in the red giant HD 122563 by placing a model \ion{Fe}{1} atom in a metal-poor
red-giant atmosphere.  She found that the \ion{Fe}{1} abundance derived from
an LTE analysis should be increased by 0.2 dex.  Unfortunately, a
comparison with Hipparcos-based gravities is not possible for giant
stars as it is for dwarfs.  For a subset of stars, we have
interpolated another set of models and chosen the \logg which made the
\ion{Fe}{1} abundance 0.2 dex smaller than the \ion{Fe}{2} abundance.  In general,
this increased the \logg by 0.4 dex over our original estimate.  Table
6 summarizes the effect on the derived abundances.

\subsubsection{NLTE effects on abundances}

We have used an LTE analysis in deriving all our abundances.  However,
for some elements, particularly Na and Al, this is inadequate.  For
\ion{Na}{1}, Gratton \etal{} (1999) calculated that an LTE analysis of
a [Fe/H]$=-2$, \teff=4000K giant using the resonance lines at 5889-5895
\AA{} underestimates the true Na abundance by $\sim$ 0.5 dex.  A
somewhat hotter 5000 K giant, on the other hand, has its Na abundance
overestimated by $\sim$0.1 dex.  Unfortunately, Mashonkina,
Shimanskii, \& Sakhibullin (2000) found very different results.  The
NLTE corrections are always negative, and start from $-0.05$ dex for a
4000 K, metal-poor giant, but reach $-0.7$ dex for a 5000 K giant.
The many differences between the two analyses, including the
NLTE code used, the number of transitions allowed, the cross-sections,
and the UV flux, are large enough that NLTE calculations of
\ion{Na}{1} have not yet converged.  We could find no similar
calculations for the \ion{Al}{1} resonance lines in metal-poor giants.
 The 3961 \AA{} resonance line, the line
used here, has a $\sim$0.5 dex correction in metal-poor dwarfs
(Baum\"uller \& Gehren 1997).  Ryan \etal{} (1996) saw no disagreement
in trends in [Al/Fe] between dwarfs and giants, which indicates that a
similarly sized NLTE effect occurs in metal-poor giants.  We have
listed LTE Na and Al abundances, but because of NLTE corrections
and our reliance on a few, very strong lines for these
elements, our Na and Al abundances have large errors
($\sim$ 0.5 dex) and that the large scatter in [Na/Fe] and [Al/Fe]
values is due to observational error.  An NLTE analysis gives results
different from an LTE analysis for other elements, such as 
\ion{Mg}{1} (e.g. Gratton \etal 1999), as well.  The corrections are smaller
than for the resonance lines of \ion{Na}{1} and \ion{Al}{1}, or are
not known at all.  Evidence for NLTE in our analysis in discussed in \S 3.4
as well.

\subsection{Consistency Checks}

To see how our spectroscopically derived
\teff{} and \logg values compare to theoretical \teff{}, \logg
relations, we plot 
our derived model atmospheres on top of isochrones from Bergbusch 
and VandenBerg (1992) and Demarque \etal{} (1996) 
(Figure 7).  Our \teff-\logg{} relation is slightly steeper than the
isochrones predict, but the overall agreement is good.  Four stars clearly
lie off of the RGB sequence; it is probable that these are AGB stars.
Indeed, those four were identified as AGB stars by Bond (1980) on the
basis of Str\"omgren photometry.

\begin{figure}[htb]
\begin{center}
\includegraphics[width=3.0in,angle=0]{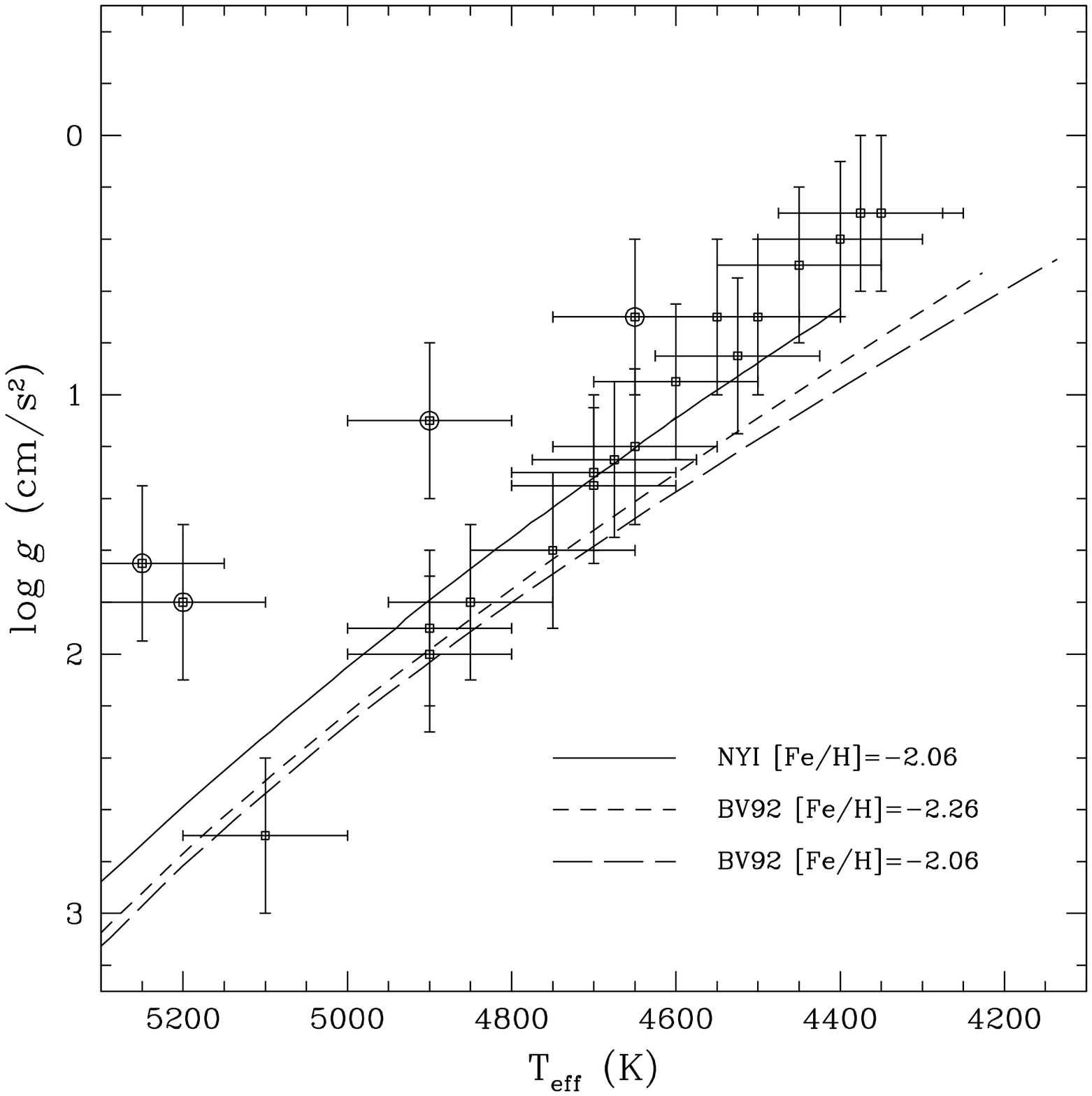}
\caption{Our model atmosphere \teff{} and \logg compared with the N
ew Yale
Isochrones (NYI) (Demarque \etal{} 1996) and Bergbusch \& VandenBerg 1992 
(BV92).
We show the lowest metallicity 15 Gyr isochrone calculated by each 
group as well as the [Fe/H]=$-2.06$ isochrone for BV92 to illustrate 
the effect of metallicity.  The four circled stars (BD -11 145, HD 108577,
BD +9 3223, and BD +17 3248) were identified as AGB stars by Bond 1980.}

\end{center}
\end{figure}

We can also derive a \logg based combining the Stefan-Boltzmann equation
with the law of gravity (e.g. Mihalas \& Binney 1982):
\begin{equation}
\rm{log}{} g= -12.50 +0.4M_{bol} + \rm{log}(M/M_{\odot}) +4 \rm{log}({\rm T_{eff}})
\end{equation}
We adopted a value of 0.8 M$_{\odot}$ for the mass of all of our stars.
To get M$_{bol}$ as a function of \teff, we began by using the stars in
the globular clusters M92 and M15 to calibrate an empirical T$_{\rm eff}$$-M_{V}$ relation.  We also needed a bolometric correction.
For M92, we used the \teff{} values and $V$ magnitudes 
from Sneden \etal{} (1991)
and the bolometric corrections from Montegriffo \etal{} (1998).  The distance
to M92 was derived using the Sandage \& Walker (1966) $V$ for the RR Lyraes
and M$_V=0.36$ for RR Lyraes at the metallicity of M92 
(Silbermann \& Smith 1995).  The resulting distance modulus of 14.65 magnitudes
agrees very well with the distance modulus of Pont \etal{} (1998) from
main-sequence fitting with subdwarfs (14.67 magnitudes).  
For M15, we took \teff{} values and M$_{bol}$ magnitudes from Sneden \etal{}
(1997).  They used a distance based on the Silbermann 
\& Smith RRLyrae magnitude and bolometric corrections from Worthey (private
communication)
to find M$_{bol}$. Armed with our spectroscopic \teff, we could then
derive M$_{bol}$ from the fiducials provided by each cluster and calculate
an evolutionary log {\it g}.  The 
\logg values found using Equation 1
are included in Table 4 as log {\it g}$_{M92}$ and log {\it g}$_{M15}$.  
We have
confined our comparison to our stars that have [Fe/H]$<-2.0$ and
\teff{} $<$ 4650 to overlap with the metallicity and temperature
ranges for the globular stars.  Our spectroscopic \logg
values are significantly lower than log {\it g}$_{M92}$; they are in
much better agreement with log {\it g}$_{M15}$.  This is
almost entirely due to the different bolometric corrections used for the
two clusters, since the Worthey BC is up to 0.5 mag brighter than those of
Montegriffo.  The \teff$-$M$_{bol}$ relation on the
giant branch is steep, and the \teff{} for the globular cluster stars
from the literature are potentially subjected to all the uncertainties
we have discussed, including the fact that they are based on MARCS
models while we used Kurucz models.    
Causes of the
offset may be normalization errors in the \ion{Fe}{1} or \ion{Fe}{2} \gf{} values or 
incorrect \gf{} values, especially for \ion{Fe}{2}.  Finally, the NLTE problems
with \ion{Fe}{1} discussed in \S 3.3.4 may be biasing our answers high.  We note
that \teff$\sim$ 4650 is when our log {\it g} values begin to lie above the
\teff$-$\logg{} relation for the theoretical isochrones.
The effect on our abundances if we forced our \logg{} to agree with
the M15 and M92 relations can be gauged by looking at the 
$\Delta$ log $\epsilon$(NLTE) in Table 6, since those abundances were calculated
with \logg{} increased by $\sim$0.4 dex.

\begin{figure}[htb]
\begin{center}
\includegraphics[width=3.0in,angle=0]{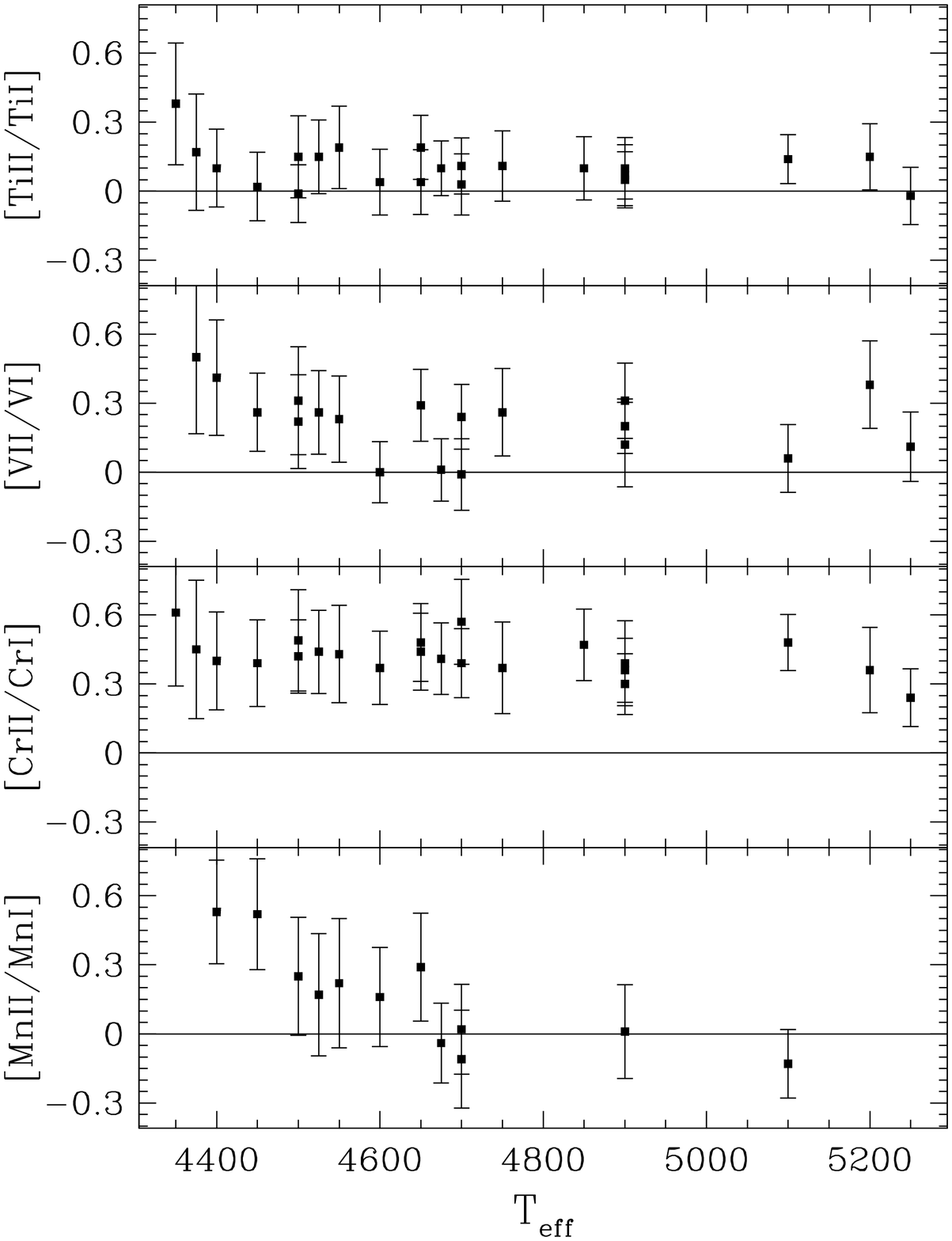}
\caption{Ratio of abundances derived from ionized and neutral lines.  The
two species give answers which often differ by 0.2 dex.  The disagreement
is worse at lower \teff{}.  The errorbars are derived using the method
discussed in \S 4.2 and reflect the random uncertainites in \teff{}, \logg{},
and $\xi$.}

\end{center}
\end{figure}

A final consistency check is comparing the abundances derived for
neutral and ionized species for elements other than Fe.  Figure 8 shows
[\ion{Ti}{2}/\ion{Ti}{1}], [\ion{V}{2}/\ion{V}{1}], [\ion{Cr}{2}/\ion{Cr}{1}], and [\ion{Mn}{2}/\ion{Mn}{1}].
  While in many
cases the abundances agree within the errors, there is a systemic
offset with the ionized species giving higher abundances.  Adoption
of either the higher gravity models from Fe NLTE corrections from \S 3.4.3
or the higher temperature models based on colors makes the discrepancy worse.
NLTE effects on the elements considered here, such as the overionization of the neutral species, could be
causing the observed discrepancy.  Unfortunately, we are not aware
of any recent calculations of NLTE effects on Ti, Cr, V and Mn in metal-poor
giants, so whether that explanation can explain the pattern seen is unclear,
especially since Ti, Cr, and V have lower ionization potentials than Mn, but the
magnitude of the discrepancy is about the same for Ti, V, and Mn, while larger
for Cr.
For each element, there are some specific possible reasons why an
offset exists.
V and Mn abundances
are affected by hyperfine splitting, which we have attempted to take into 
account,
but may still cause some of the offset (see \S 4.1).  
The normalization problems
of Ti and \ion{Cr}{2} have already been discussed.  The \ion{Ti}{2} values used in our study give
a solar Ti abundance 0.06 dex higher than the \ion{Ti}{1} we used, and both sets
of \gf{} values gave a higher solar Ti abundance than the meteoritic abundance.
  Finally, the \ion{Mn}{2} lines are at 3400
\AA{}, where both continuum placement and unknown blending may cause
problems.  Agreement between neutral and ionized species is a stringent
test of the atmospheric models, and clearly the current situation is
unsatisfactory.

\subsection{Comparison with Previous Work}

In Table 7, we list the model atmosphere parameters
and resulting [Fe/H] from a selection of previous studies.
All the
studies determined \logg from the ionization balance of \ion{Fe}{1} and \ion{Fe}{2}, 
sometimes including \ion{Ti}{1} and \ion{Ti}{2} as well.  $\xi$ was found by the standard method of
having no trend in abundance with logRW.  The main difference in technique
between the studies was in the \teff{} determination.  Some studies used
the spectroscopic method of eliminating trends in abundance versus 
E.P. for \ion{Fe}{1} lines, while others relied on photometric
colors.  We have noted the method used in Table 7
as either Spec. or Phot.  Our temperatures and gravities are in general
lower than previous determinations, particularly the photometric
determinations, which agrees with our discussion in \S 3.3.1.  However,
the agreement improves when we consider only the more recent
determinations.  The 400 K offset 
between our temperature   
McWilliam \etal's for HD 128279 is due to the large reddening adopted
for that star by McWilliam \etal{} (1995b).  

\begin{deluxetable}{llccccl}
\tablewidth{0pt}
\tablenum{7}
\tablecaption{Comparison with Previous Model Atmosphere Parameters}
\tablehead{
\colhead{Star}  & \colhead{Study} & \colhead{\teff} & \colhead{logg}
& \colhead{$\xi$} & \colhead{[Fe/H]} &\colhead {\teff} \\
\colhead{} & \colhead{} & \colhead{K} & \colhead {gm cm$^{-2}$} & 
\colhead{km/s} & \colhead{} & \colhead{method} }
\startdata
HD 29574 & this study & 4350 & 0.30 & 2.30 & $-1.86$ \\
 & Shetrone 1996 & 4100 & 0.00 & 1.60 & $-1.93$ & Spec.\\
HD 63791 & this study & 4750 & 1.60 & 1.70 & $-1.73$ & \\ 
 & Gilroy \etal 1988 & 4800 & 1.90 & 2.20 & $-1.72$ & Spec. \\
HD 88609 & this study & 4400 & 0.40 & 2.40 &  $-2.97$ & \\
 & Luck \& Bond 1985  & 4500 & 0.80  & 3.20 & $-2.66$ & Spec. \\
 & Gilroy \etal 1988 & 4500 & 1.10 & 2.80 & $-2.78$  & Spec. \\
HD 115444 & this study & 4500 & 0.70 & 2.25 & $-3.15$ & \\
 & Gilroy \etal 1988 & 4800 & 2.00 & 2.20 & $-2.64$ & Spec. \\
 & Westin \etal 1999 & 4650 & 1.50 & 2.10 & $-2.90$ & Spec. \\
HD 122563 & this study & 4450 & 0.50 & 2.30 & $-2.76$ & \\
 & Luck \& Bond 1985 & 4600 & 1.40 & 2.80  & $-2.35$ & Spec. \\
 & Gilroy \etal 1988 & 4600 & 1.20 & 2.30 & $-2.45$ & Spec. \\
 & Peterson \etal 1990 & 4500 & 0.75 & 2.50 & $-2.93$ & Spec. \\
 & Gratton \& Sneden 1994 & 4590 & 1.17 & 2.30 & $-2.81$ & Phot. \\
 & Ryan \etal 1996 & 4650 & 1.40 & 2.60 & $-2.68$ & Phot. \\
 & Westin \etal 1999 & 4500 & 1.30 & 2.50 & $-2.74$ & Spec. \\
HD 126587 & this study & 4675 & 1.25 & 1.90 & $-3.08$ \\ 
 & Luck \& Bond 1985 &  4750 & 1.10 & 2.00 & $-2.66$ & Spec. \\
 & McWilliam \etal 1995b & 4910 & 1.85 & 2.03 & $-2.85$ & Phot. \\
HD 128279 & this study & 5100 & 2.70 & 1.40 & $-2.39$  & \\
 & Gilroy \etal 1988 & 5000 & 2.20 & 1.40 & $-2.21$ & Spec. \\
 & Peterson \etal 1990 & 5125 & 2.20 & 2.00 & $-2.50$ & Spec. \\
 & McWilliam \etal 1995b & 5480 & 3.10 & 1.98 & $-2.08$ & Phot. \\
HD 165195 & this study & 4375 & 0.30 & 2.50 & $-2.32$ &  \\
 & Gilroy \etal 1988 & 4500 & 1.50 & 2.80 & $-2.23$ & Spec. \\
 & Gratton \& Sneden 1994 & 4507 & 1.45 & 3.20 & $-2.25$ & Phot. \\
HD 186478 & this study & 4525 & 0.85 & 2.00 & $-2.61$ & \\
 & McWilliam \etal 1995b & 4650 & 0.95 & 2.71 & $-2.58$ & Phot. \\
HD 216143 & this study & 4500 & 0.70 & 2.10 & $-2.23$ & \\
 & Shetrone 1996 & 4400 & 0.70 & 1.80 & $-2.26$ & Spec.\\
BD -18 5550 & this study &  4600 & 0.95 & 1.90 & $-3.05$ & \\
 & Barbuy \etal 1985 & 4580 & 1.00 & 1.50 & $-3.05$ & Phot. \\
 & Luck \& Bond 1985 & 4750 & 0.80 & 2.90 & $-2.66$ & Spec. \\
 & Gilroy \etal 1988 & 4600 & 1.30 & 3.00 & $-2.92$ & Spec \\
 & Peterson \etal 1990 & 4750  & 1.12 &  2.50 & $-2.90$ & Spec. \\
 & McWilliam \etal 1995b & 4790 & 1.15 & 2.14 & $-2.91$ & Phot. \\
BD +4 2621 & this study & 4650 & 1.20 & 1.80 & $-2.52$ & \\
 & Luck \& Bond 1985 & 4750 & 1.10 & 2.00 & $-2.22$ & Spec. \\
M92VII-18 & this study & 4250 & 0.20 & 2.30 & $-2.27$ & \\
 & Shetrone 1996 & 4230 & 0.70 & 2.50 & $-2.33$ & Spec. \\
\enddata
\end{deluxetable}

\section{Abundances}
Abundances of 30 elements are listed in Table 8.  For some rarely measured
elements, such as Os, we include generous upper limits derived from
spectral synthesis to help constrain the heavy-element abundance ratios
in metal-poor stars (see discussion in Johnson \& Bolte 2001).   
To derive abundances, we used the program MOOG.  
MOOG calculates abundances and synthetic spectra based
on input model atmospheres and the assumption of LTE.  For those 
elements unaffected by hyperfine splitting (HFS), we
used the routine {\it abfind} in MOOG to calculate abundances
from our EWs.  For lines affected by HFS, we used the routine {\it blends}, 
which allowed us to derive abundances from EWs while considering
 multiple lines for one element.  Some elements had only blended lines.
We synthesized these spectral regions using MOOG, and linelists
from Sneden \etal{} (1996).  They are noted in Table 2 as syn.
We used the Uns\"old (1955) approximation for calculating the damping constants. 
The solar abundances adopted when needed were the photospheric abundances from
Anders \& Grevesse (1989), except for Fe.  We used log$\epsilon_{Fe}$=7.52
(Bi\'emont \etal{} 1991).  The last column of Table 8 lists the solar log$\epsilon$ for reference.
\begin{deluxetable}{lrcccrcccrccc}
\tablenum{8A}
\tablewidth{0pt}
\tablecaption{Abundances}
\tablehead{\colhead{Element} &\multicolumn{4}{c}{HD29574} & 
\multicolumn{4}{c}{HD63791} & \multicolumn{4}{c}{HD88609}    \\
\colhead{} & \multicolumn{1}{r}{log $\epsilon$} & \colhead{$\sigma$} 
& \colhead{$\sigma_{tot}$} & \colhead{N$_{lines}$} & \multicolumn{1}{r}{log $\epsilon$} 
& \colhead{$\sigma$} 
& \colhead{$\sigma_{tot}$} & \colhead{N$_{lines}$} & \multicolumn{1}{r}{log $\epsilon$} 
& \colhead{$\sigma$} 
& \colhead{$\sigma_{tot}$} & \colhead{N$_{lines}$} }
\startdata
NaI  &   \nodata & \nodata & \nodata & \nodata & \phs$4.49$ & 0.10 & 0.21 &   2 &   \phs $3.36$ & 0.10 & 0.22 &   2 \\
MgI  & \phs$6.62$ & 0.11 & 0.21 &   3 & \phs$ 6.59$ & 0.15 & 0.16 &   5 & \phs
$5.44$ & 0.16 & 0.19 &   8 \\
AlI  &   \phs$4.21$ & 0.10 & 0.26 &   1 &    \nodata & \nodata & \nodata & \nodata & \phs$ 3.53$ & 0.10 & 0.31 &   1 \\
SiI  &\phs$ 6.11$ & 0.15 & 0.13 &   3 & \phs$ 6.23$ & 0.12 & 0.11 &   4 & \phs$ 5.46$ & 0.20 & 0.29 &   1 \\
CaI  &\phs$ 4.86$ & 0.12 & 0.16 &  25 & \phs$ 5.09$ & 0.07 & 0.14 &  25 & \phs$ 3.77$ & 0.09 & 0.07 &  18 \\
ScII &\phs$ 1.35$ & 0.09 & 0.12 &   8 & \phs$ 1.56$ & 0.06 & 0.16 &  10 & \phs$ 0.20$ & 0.09 & 0.06 &  12 \\
TiI  &\phs$ 3.20$ & 0.15 & 0.26 &  39 & \phs$ 3.53$ & 0.10 & 0.18 &  40 & \phs$ 2.21$ & 0.15 & 0.15 &  29 \\
TiII &\phs$ 3.58$ & 0.20 & 0.14 &  16 & \phs$ 3.64$ & 0.09 & 0.16 &  18 & \phs$ 2.31$ & 0.08 & 0.06 &  18 \\
VI   &   \phs$1.91$ & 0.01 & 0.27 &   2 & \phs$ 2.27$ & 0.02 & 0.17 &   2 &    \phs $0.89$ & 0.02 & 0.14 &   3 \\
VII  &   \nodata & \nodata & \nodata & \nodata & \phs$ 2.53$ & 0.10 & 0.15 &   1 & \phs$ 1.30$ & 0.12 & 0.18 &   3 \\
CrI  &   $3.58$ & 0.09 & 0.26 &  14 &  \phs  $3.87$ & 0.07 & 0.20 &  14 &   \phs $2.44$ & 0.06 & 0.12 &  13 \\
CrII &\phs$ 4.19$ & 0.05 & 0.09 &   1 & \phs$ 4.24$ & 0.06 & 0.12 &   2 & \phs$ 2.84$ & 0.13 & 0.12 &   2 \\
MnI  &  \phs $3.32$ & 0.15 & 0.18 &   5 &    \phs $3.32$& 0.06 & 0.13 & 6 &  \phs $2.11$ & 0.18 & 0.14 &   5 \\
MnII &   \nodata & \nodata & \nodata & \nodata &    \nodata & \nodata & \nodata & \nodata & \phs$ 2.64$ & 0.11 & 0.22 &   2 \\
FeI  &  \phs $5.64$ & 0.16 & 0.22 & 151 &   \phs $5.80$ & 0.16 & 0.21 & 171 &   \phs $4.55$ & 0.18 & 0.16 & 156 \\
FeII &  \phs $5.68$ & 0.13 & 0.12 &  15 &   \phs $5.78$ & 0.14 & 0.16 &  24 &   \phs $4.56$ & 0.10 & 0.07 &  18 \\
CoI  &\phs$ 3.14$ & 0.40 & 0.33 &   3 & \phs$ 3.49$ & 0.43 & 0.32 &   4 & \phs$ 2.29$ & 0.13 & 0.19 &   7 \\
NiI  &   \phs$4.33$ & 0.16 & 0.15 &  34 & \phs $4.50$ & 0.16 & 0.13 &  39 & \phs$ 3.35$ & 0.23 & 0.25 &  23 \\
ZnI  &\phs$ 2.74$ & 0.04 & 0.08 &   1 & \phs$ 2.93$ & 0.04 & 0.10 &   2 & \phs$ 1.88$ & 0.04 & 0.03 &   2 \\
YII  &  \phs $0.27$ & 0.10 & 0.16 &   6 & \phs$0.28$ & 0.05 & 0.17 &   5 &    $-0.80$ & 0.21 & 0.09 &  13 \\
ZrII &\phs$ 1.07$ & 0.21 & 0.16 &   3 & \phs$ 1.01$ & 0.13 & 0.15 &   4 & \phs$ 0.02$ & 0.11 & 0.09 &  11 \\
BaII &\phs$ 0.54$ & 0.11 & 0.26 &   4 & \phs$ 0.42$ & 0.07 & 0.26 &   4 &    $-1.92$ & 0.05 & 0.10 &   4 \\
LaII &   $-0.82$ & 0.08 & 0.08 &   4 &    $-0.73$ & 0.05 & 0.14 &   4 &    \nodata & \nodata & \nodata & \nodata \\
CeII & $-0.22$ & 0.02 & 0.11 &   2 &    $-0.26$ & 0.08 & 0.14 &   5 &    \nodata & \nodata & \nodata & \nodata \\
PrII &   \nodata & \nodata & \nodata & \nodata &    \nodata & \nodata & \nodata & \nodata & $<$$-1.26$ & \nodata & \nodata & \nodata \\
NdII &$ -0.07$ & 0.21 & 0.12 &  12 & $-0.12$ & 0.29 & 0.17 &   9 &    $-1.93$ & 0.20 & 0.20 &   1 \\
SmII &$ -0.39$ & 0.25 & 0.13 &   7 & $ -0.47$ & 0.08 & 0.13 &   5 &    \nodata & \nodata & \nodata & \nodata \\
EuII &$ -1.15$ & 0.20 & 0.20 &   1 & $ -1.12$ & 0.20 & 0.24 &   1 &    $-2.96$ & 0.20 & 0.20 &   1 \\
GdII &   \nodata & \nodata & \nodata & \nodata &    \nodata & \nodata & \nodata & \nodata &    \nodata & \nodata & \nodata & \nodata \\
TbII &   \nodata & \nodata & \nodata & \nodata & $<$$-0.58$ & \nodata & \nodata & \nodata & $<$$ -1.72$ & \nodata & \nodata & \nodata \\
DyII &   \nodata & \nodata & \nodata & \nodata &    \nodata & \nodata & \nodata & \nodata &    \nodata & \nodata & \nodata & \nodata \\
HoII &   \nodata & \nodata & \nodata & \nodata & $<$$ -0.23$ & \nodata & \nodata & \nodata & $<$$ -2.17$ & \nodata & \nodata & \nodata \\
ErII &   \nodata & \nodata & \nodata & \nodata &    \nodata & \nodata & \nodata & \nodata &    $-2.93$ & 0.10 & 0.11 &   1 \\
TmII &   \nodata & \nodata & \nodata & \nodata &    \nodata & \nodata & \nodata & \nodata & $<$$ -2.14$ & \nodata & \nodata & \nodata \\
YbII &   \nodata & \nodata & \nodata & \nodata &    \nodata & \nodata & \nodata & \nodata &    $-2.89$ & 0.20 & 0.21 &   1 \\
HfII &   \nodata & \nodata & \nodata & \nodata &    \nodata & \nodata & \nodata & \nodata &    \nodata & \nodata & \nodata & \nodata \\
OsI  &$<$$ 0.19$ & \nodata & \nodata & \nodata & $<$$ 0.62$ & \nodata & \nodata & \nodata & $<$$ -0.62$ & \nodata & \nodata & \nodata \\
\enddata
\end{deluxetable}

\begin{deluxetable}{lrcccrcccrccc}
\tablenum{8B}
\tablewidth{0pt}
\tablecaption{Abundances}
\tablehead{\colhead{Element} &\multicolumn{4}{c}{HD 108577} & 
\multicolumn{4}{c}{HD 115444} & \multicolumn{4}{c}{HD 122563}    \\
\colhead{} & \multicolumn{1}{r}{log $\epsilon$} & \colhead{$\sigma$} 
& \colhead{$\sigma_{tot}$} & \colhead{N$_{lines}$} & \multicolumn{1}{r}{log $\epsilon$} 
& \colhead{$\sigma$} 
& \colhead{$\sigma_{tot}$} & \colhead{N$_{lines}$} & \multicolumn{1}{r}{log $\epsilon$} 
& \colhead{$\sigma$} 
& \colhead{$\sigma_{tot}$} & \colhead{N$_{lines}$} }
\startdata
NaI  &\phs$ 4.19$ & 0.10 & 0.21 &   2 & \phs$3.57$ & 0.10 & 0.20 &   2 & \phs$ 3.70$ & 0.22 & 0.28 &   2 \\
MgI  &\phs$ 5.86$ & 0.05 & 0.12 &   5 & \phs$5.26$ & 0.10 & 0.16 &   8 & \phs$ 5.52$ & 0.12 & 0.15 &   5 \\
AlI  &   $3.69$ & 0.10 & 0.25 &   1 &    $3.24$ & 0.10 & 0.28 &   1 &    $3.61$ & 0.10 & 0.28 &   1 \\
SiI  &\phs$ 5.99$ & 0.20 & 0.26 &   1 & \phs$5.00$ & 0.20 & 0.24 &   1 & \phs$5.50$ & 0.20 & 0.28 &   1 \\
CaI  &\phs$ 4.37$ & 0.06 & 0.07 &  21 & \phs$ 3.63$ & 0.06 & 0.06 &  12 & \phs$ 3.90$ & 0.07 & 0.07 &  20 \\
ScII &\phs$ 0.81$ & 0.09 & 0.14 &  11 & \phs$ 0.02$ & 0.09 & 0.08 &   9 & \phs$ 0.43$ & 0.10 & 0.08 &  10 \\
TiI  &\phs$ 2.82$ & 0.09 & 0.11 &  29 & \phs$ 2.16$ & 0.09 & 0.10 &  30 & \phs$ 2.38$ & 0.13 & 0.14 &  35 \\
TiII &\phs$ 2.87$ & 0.05 & 0.13 &  20 & \phs$ 2.15$ & 0.08 & 0.08 &  18 & \phs$ 2.40$ & 0.08 & 0.07 &  19 \\
VI   &   $1.55$ & 0.04 & 0.12 &   4 &    $0.79$ & 0.03 & 0.10 &   2 &    $1.19$ & 0.02 & 0.14 &   4 \\
VII  &\phs$ 1.75$ & 0.03 & 0.10 &   3 & \phs$ 1.01$ & 0.08 & 0.15 &   5 & \phs$ 1.45$ & 0.05 & 0.07 &   2 \\
CrI  &   $3.19$ & 0.05 & 0.11 &  12 &    $2.35$ & 0.08 & 0.10 &  12 &    $2.66$ & 0.06 & 0.12 &  14 \\
CrII &\phs$ 3.55$ & 0.05 & 0.10 &   2 & \phs$ 2.77$ & 0.05 & 0.08 &   1 & \phs$ 3.05$ & 0.06 & 0.08 &   2 \\
MnI  &   $2.83$ & 0.11 & 0.09 &   6 &    $1.92$ & 0.02 & 0.09 &   2 &    $2.43$ & 0.08 & 0.11 &   5 \\
MnII &   $2.84$ & 0.05 & 0.17 &   1 &    $2.17$ & 0.07 & 0.29 &   3 & \phs$2.95$ & 0.15 & 0.25 &   1 \\
FeI  &   $5.14$ & 0.12 & 0.13 & 168 &    $4.37$ & 0.13 & 0.11 & 149 &    $4.77$ & 0.16 & 0.15 & 161 \\
FeII &   $5.13$ & 0.10 & 0.11 &  23 &    $4.36$ & 0.08 & 0.06 &  19 &    $4.75$ & 0.11 & 0.08 &  21 \\
CoI  &\phs$ 2.58$ & 0.36 & 0.18 &   8 & \phs$ 2.13$ & 0.13 & 0.13 &   6 & \phs$2.52$ & 0.11 & 0.18 &   7 \\
NiI  &\phs$ 3.93$ & 0.13 & 0.16 &  25 & \phs$ 3.24$ & 0.18 & 0.20 &  19 & \phs$ 3.60$ & 0.14 & 0.17 &  21 \\
ZnI  &\phs$ 2.56$ & 0.04 & 0.07 &   2 & \phs$ 1.72$ & 0.04 & 0.03 &   2 & \phs$ 2.00$ & 0.04 & 0.04 &   2 \\
YII  &   $-0.52$ & 0.07 & 0.13 &  12 &    $-1.00$ & 0.10 & 0.09 &  11 &    $-0.80$ & 0.12 & 0.10 &  11 \\
ZrII &   $0.18$ & 0.08 & 0.10 &  10 & $ -0.30$ & 0.13 & 0.08 &   9 & $-0.14$ & 0.12 & 0.08 &  10 \\
BaII &   $-0.35$ & 0.10 & 0.21 &   4 &    $-1.10$ & 0.08 & 0.15 &   4 &    $-1.80$ & 0.01 & 0.11 &   4 \\
LaII &   $-1.24$ & 0.09 & 0.13 &   4 & $-1.68$ & 0.05 & 0.07 &   4 &    $-2.44$ & 0.10 & 0.11 &   1 \\
CeII &   $-1.05$ & 0.07 & 0.14 &   3 & $-1.53$ & 0.11 & 0.11 &   3 &    \nodata & \nodata & \nodata & \nodata \\
PrII &$<$$-1.17$ & \nodata & \nodata & \nodata & $-2.15$ & 0.20 & 0.05 &   2 & $<$$ -1.55$ & \nodata & \nodata & \nodata \\
NdII &\phs$ -0.82$ & 0.16 & 0.14 &   7 & $ -1.36$ & 0.20 & 0.10 &   8 &    $-1.87$ & 0.20 & 0.20 &   1 \\
SmII &$-1.15$ & 0.14 & 0.14 &   8 & $-1.59$ & 0.09 & 0.10 &   7 &    \nodata & \nodata & \nodata & \nodata \\
EuII &$-1.48$ & 0.02 & 0.12 &   2 & $ -1.82$ & 0.03 & 0.05 &   3 &    $-2.85$ & 0.20 & 0.21 &   1 \\
GdII &$ -1.11$ & 0.20 & 0.17 &   2 & $ -1.47$ & 0.20 & 0.20 &   1 &    \nodata & \nodata & \nodata & \nodata \\
TbII &$<$$ -1.43$ & \nodata & \nodata & \nodata & $ -2.41$ & 0.15 & 0.12 &   2 & $<$$ -1.81$ & \nodata & \nodata & \nodata \\
DyII &$ -0.99$ & 0.13 & 0.11 &   6 & $ -1.30$ & 0.13 & 0.07 &   9 &    \nodata & \nodata & \nodata & \nodata \\
HoII &$<$$ -0.88$ & \nodata & \nodata & \nodata &    \nodata & \nodata & \nodata & \nodata & $<$$-1.66$ & \nodata & \nodata & \nodata \\
ErII &$ -1.09$ & 0.02 & 0.11 &   3 & $ -1.40$ & 0.04 & 0.04 &   3 &    $-2.43$ & 0.10 & 0.10 &   1 \\
TmII &$-1.98$ & 0.15 & 0.19 &   1 & $ -2.36$ & 0.15 & 0.11 &   2 & $<$$-2.33$ & \nodata & \nodata & \nodata \\
YbII &\phs$-1.10$ & 0.20 & 0.28 &   1 & $ -1.28$ & 0.20 & 0.28 &   1 &    $-2.78$ & 0.20 & 0.20 &   1 \\
HfII &   \nodata & \nodata & \nodata & \nodata &    \nodata & \nodata & \nodata & \nodata & $<$$ -0.88$ & \nodata & \nodata & \nodata \\
OsI  &$<$$ -0.03$ & \nodata & \nodata & \nodata & $<$$ -0.80$ & \nodata & \nodata & \nodata & $<$$ -0.41$ & \nodata & \nodata & \nodata \\
\enddata
\end{deluxetable}

\begin{deluxetable}{lrcccrcccrccc}
\tablenum{8C}
\tablewidth{0pt}
\tablecaption{Abundances}
\tablehead{\colhead{Element} &\multicolumn{4}{c}{HD 126587} & 
\multicolumn{4}{c}{HD 128279} & \multicolumn{4}{c}{HD 165195}    \\
\colhead{} & \multicolumn{1}{r}{log $\epsilon$} & \colhead{$\sigma$} 
& \colhead{$\sigma_{tot}$} & \colhead{N$_{lines}$} & \multicolumn{1}{r}{log $\epsilon$} 
& \colhead{$\sigma$} 
& \colhead{$\sigma_{tot}$} & \colhead{N$_{lines}$} & \multicolumn{1}{r}{log $\epsilon$} 
& \colhead{$\sigma$} 
& \colhead{$\sigma_{tot}$} & \colhead{N$_{lines}$} }
\startdata
NaI  &\phs$3.61$ & 0.10 & 0.19 &   2 &    $3.83$ & 0.10 & 0.16 &   2 &    \nodata & \nodata & \nodata & \nodata \\
MgI  &\phs$ 5.19$ & 0.10 & 0.13 &   8 & $5.79 $ & 0.09 & 0.11 &   5 & \phs$ 6.02$ & 0.19 & 0.20 &   4 \\
AlI  &   $3.02$ & 0.10 & 0.24 &   1 &    $3.42$ & 0.10 & 0.21 &   1 &    $3.42$ & 0.10 & 0.35 &   1 \\
SiI  &$5.02$ & 0.20 & 0.22 &   1 & \phs$5.55$ & 0.20 & 0.23 &   1 & \phs$ 5.74$ & 0.20 & 0.18 &   2 \\
CaI  &\phs$3.74$ & 0.07 & 0.06 &  14 & \phs$4.42$ & 0.11 & 0.08 &  21 & \phs$ 4.45$ & 0.09 & 0.12 &  25 \\
ScII &\phs$ 0.12$ & 0.11 & 0.12 &   8 & \phs$0.94$ & 0.11 & 0.15 &   9 & \phs$ 0.92$ & 0.13 & 0.09 &  11 \\
TiI  &\phs$ 2.14$ & 0.09 & 0.11 &  23 & \phs$ 2.82$ & 0.11 & 0.13 &  25 & \phs$ 2.86$ & 0.12 & 0.23 &  37 \\
TiII &\phs$ 2.24$ & 0.06 & 0.11 &  15 & \phs$ 2.96$ & 0.03 & 0.15 &  16 & \phs$ 3.03$ & 0.08 & 0.10 &  17 \\
VI   &   $0.85$ & 0.01 & 0.12 &   2 & \phs$ 1.61$ & 0.07 & 0.13 &   2 &    $1.59$ & 0.11 & 0.26 &   3 \\
VII  &   $0.86$ & 0.13 & 0.08 &   5 & \phs$ 1.67$ & 0.23 & 0.16 &   6 & \phs$2.09$ & 0.10 & 0.12 &   1 \\
CrI  &   $2.38$ & 0.11 & 0.12 &   9 &    $3.18$ & 0.07 & 0.11 &  12 &    $3.21$ & 0.08 & 0.21 &  14 \\
CrII &\phs$2.79$ & 0.05 & 0.09 &   1 & \phs$ 3.66$ & 0.03 & 0.10 &   2 & \phs$ 3.66$ & 0.05 & 0.12 &   1 \\
MnI  &   $1.98$ & 0.08 & 0.11 &   1 &    $2.70$ & 0.05 & 0.09 &   3 &    $2.84$ & 0.14 & 0.14 &   6 \\
MnII &   $1.94$ & 0.03 & 0.12 &   3 & $2.57$ & 0.05 & 0.09 & 3 &    \nodata & \nodata & \nodata & \nodata \\
FeI  &   $4.44$ & 0.09 & 0.12 & 137 &    $5.12$ & 0.12 & 0.14 & 147 &    $5.20$ & 0.18 & 0.19 & 163 \\
FeII &   $4.44$ & 0.06 & 0.09 &  18 &    $5.14$ & 0.11 & 0.12 &  20 &    $5.20$ & 0.14 & 0.10 &  23 \\
CoI  &\phs$2.09$ & 0.13 & 0.13 &   7 & \phs$ 2.64$ & 0.12 & 0.14 &   7 & \phs$ 2.84$ & 0.26 & 0.27 &   3 \\
NiI  &\phs$ 3.20$ & 0.12 & 0.18 &  17 &    $3.84$ & 0.13 & 0.15 &  16 & \phs$ 3.97$ & 0.17 & 0.14 &  33 \\
ZnI  &   \nodata & \nodata & \nodata & \nodata &    \nodata & \nodata & \nodata & \nodata & \phs$2.30$ & 0.04 & 0.04 &   2 \\
YII  &   $-1.07$ & 0.07 & 0.09 &  11 &    $-0.72$ & 0.09 & 0.15 &   6 &    $-0.38$ & 0.11 & 0.08 &   6 \\
ZrII &\phs$ -0.36$ & 0.07 & 0.08 &   6 &    $-0.09$ & 0.06 & 0.14 &   5 & \phs$ 0.46$ & 0.20 & 0.11 &   4 \\
BaII &   $-1.08$ & 0.12 & 0.16 &   4 &    $-0.74$ & 0.03 & 0.17 &   3 &    $-0.43$ & 0.04 & 0.20 &   4 \\
LaII &   $-1.90$ & 0.08 & 0.12 &   3 &    $-1.47$ & 0.15 & 0.16 &   4 &    $-1.25$ & 0.05 & 0.04 &   4 \\
CeII &   \nodata & \nodata & \nodata & \nodata &    \nodata & \nodata & \nodata & \nodata &    $-0.88$ & 0.08 & 0.08 &   3 \\
PrII &$<$$-1.37$ & \nodata & \nodata & \nodata & $<$$ -0.68$ & \nodata & \nodata & \nodata &    \nodata & \nodata & \nodata & \nodata \\
NdII &$-1.37 $ & 0.20 & 0.18 &   2 & $ -0.79$ & 0.20 & 0.24 &   1 & $ -0.74$ & 0.20 & 0.09 &  12 \\
SmII &   \nodata & \nodata & \nodata & \nodata &    \nodata & \nodata & \nodata & \nodata & $ -1.05$ & 0.14 & 0.08 &   7 \\
EuII &$-2.15$ & 0.20 & 0.23 &   1 & $-1.78 $ & 0.20 & 0.25 &   1 & $ -1.32$ & 0.20 & 0.20 &   1 \\
GdII &$-1.72 $ & 0.20 & 0.18 &   2 &    \nodata & \nodata & \nodata & \nodata &    \nodata & \nodata & \nodata & \nodata \\
TbII &$<$$ -1.93$ & \nodata & \nodata & \nodata & $<$$ -1.24$ & \nodata & \nodata & \nodata &    \nodata & \nodata & \nodata & \nodata \\
DyII &$-1.56 $ & 0.20 & 0.13 &   3 & $ -1.02$ & 0.20 & 0.20 &   2 & $ -0.56$ & 0.20 & 0.15 &   2 \\
HoII &   \nodata & \nodata & \nodata & \nodata & $<$$ -0.89$ & \nodata & \nodata & \nodata &    \nodata & \nodata & \nodata & \nodata \\
ErII &$-1.71$ & 0.10 & 0.12 &   2 & $ -1.26$ & 0.10 & 0.17 &   1 &    \nodata & \nodata & \nodata & \nodata \\
TmII &$<$$ -2.05$ & \nodata & \nodata & \nodata & $<$$-1.26$ & \nodata & \nodata & \nodata &    \nodata & \nodata & \nodata & \nodata \\
YbII &$ -1.85$ & 0.20 & 0.22 &   1 &    $-1.61$ & 0.20 & 0.25 &   1 &    \nodata & \nodata & \nodata & \nodata \\
HfII &   \nodata & \nodata & \nodata & \nodata &    \nodata & \nodata & \nodata & \nodata &    \nodata & \nodata & \nodata & \nodata \\
OsI  &$<$$ -0.73$ & \nodata & \nodata & \nodata & $<$$ 0.06$ & \nodata & \nodata & \nodata & $<$$ -0.27$ & \nodata & \nodata & \nodata \\
\enddata
\end{deluxetable}

\begin{deluxetable}{lrcccrcccrccc}
\tablenum{8D}
\tablewidth{0pt}
\tablecaption{Abundances}
\tablehead{\colhead{Element} &\multicolumn{4}{c}{HD 186478} & 
\multicolumn{4}{c}{HD 216143} & \multicolumn{4}{c}{HD 218857}    \\
\colhead{} & \multicolumn{1}{r}{log $\epsilon$} & \colhead{$\sigma$} 
& \colhead{$\sigma_{tot}$} & \colhead{N$_{lines}$} & \multicolumn{1}{r}{log $\epsilon$} 
& \colhead{$\sigma$} 
& \colhead{$\sigma_{tot}$} & \colhead{N$_{lines}$} & \multicolumn{1}{r}{log $\epsilon$} 
& \colhead{$\sigma$} 
& \colhead{$\sigma_{tot}$} & \colhead{N$_{lines}$} }
\startdata
NaI  &$ 3.88$ & 0.19 & 0.28 &   2 &    $3.94$ & 0.10 & 0.29 &   2 & $ 4.14$ & 0.10 & 0.19 &   2 \\
MgI  &$5.71$ & 0.17 & 0.16 &   5 & $ 6.04$ & 0.24 & 0.19 &   5 & $ 6.07$ & 0.11 & 0.13 &   3 \\
AlI  &   $3.63$ & 0.10 & 0.28 &   1 &    $3.90$ & 0.10 & 0.28 &   1 &    $3.23$ & 0.10 & 0.25 &   1 \\
SiI  &$ 5.65$ & 0.16 & 0.10 &   4 & $ 5.73$ & 0.08 & 0.08 &   4 & $ 5.63$ & 0.28 & 0.18 &   3 \\
CaI  &$ 4.19$ & 0.04 & 0.08 &  23 & $ 4.42$ & 0.06 & 0.10 &  25 & $4.56$ & 0.09 & 0.10 &  22 \\
ScII &$ 0.62$ & 0.11 & 0.11 &  10 & $ 0.96$ & 0.15 & 0.12 &   7 &    $0.86$ & 0.09 & 0.15 &   9 \\
TiI  &$ 2.56$ & 0.10 & 0.15 &  41 & $ 2.87$ & 0.13 & 0.18 &  39 & $2.91$ & 0.08 & 0.14 &  27 \\
TiII &$ 2.71$ & 0.08 & 0.11 &  19 & $ 3.02$ & 0.09 & 0.12 &  18 & $3.01$ & 0.07 & 0.16 &  16 \\
VI   &   $1.27$ & 0.06 & 0.15 &   6 &    $1.65$ & 0.04 & 0.19 &   4 &    $1.64$ & 0.05 & 0.15 &   1 \\
VII  &$ 1.53$ & 0.11 & 0.07 &   3 & $ 1.96$ & 0.10 & 0.11 &   1 &    \nodata & \nodata & \nodata & \nodata \\
CrI  &   $2.86$ & 0.05 & 0.14 &  13 &    $3.32$ & 0.05 & 0.17 &  14 &    $3.39$ & 0.06 & 0.13 &  13 \\
CrII &$ 3.30$ & 0.03 & 0.07 &   2 & $ 3.81$ & 0.02 & 0.08 &   2 & $3.86$ & 0.07 & 0.12 &   2 \\
MnI  &   $2.47$ & 0.13 & 0.11 &   7 &    $2.99$ & 0.13 & 0.13 &   6 &    $2.85$ & 0.05 & 0.11 &   4 \\
MnII &   $2.64$ & 0.02 & 0.22 &   2 &    \nodata & \nodata & \nodata & \nodata &    \nodata & \nodata & \nodata & \nodata \\
FeI  &   $4.91$ & 0.12 & 0.16 & 167 &    $5.29$ & 0.15 & 0.19 & 165 &    $5.33$ & 0.11 & 0.16 & 148 \\
FeII &   $4.92$ & 0.10 & 0.08 &  24 &    $5.28$ & 0.11 & 0.10 &  25 &    $5.33$ & 0.13 & 0.14 &  21 \\
CoI  &$ 2.48$ & 0.15 & 0.19 &   5 & $ 2.95$ & 0.22 & 0.25 &   4 &    $2.69$ & 0.11 & 0.18 &   2 \\
NiI  &   $3.56$ & 0.20 & 0.17 &  30 & $ 4.04$ & 0.17 & 0.13 &  32 &    $4.01$ & 0.19 & 0.12 &  17 \\
ZnI  &$ 2.20$ & 0.04 & 0.05 &   1 & $ 2.53$ & 0.04 & 0.05 &   2 & $ 2.60$ & 0.04 & 0.09 &   2 \\
YII  &   $-0.48$ & 0.07 & 0.12 &  14 &    $-0.16$ & 0.18 & 0.14 &   6 &    $-0.43$ & 0.06 & 0.13 &   4 \\
ZrII &$ 0.29$ & 0.08 & 0.06 &  11 & $ 0.53$ & 0.15 & 0.10 &   4 &    \nodata & \nodata & \nodata & \nodata \\
BaII &   $-0.55$ & 0.15 & 0.22 &   4 &    $-0.30$ & 0.09 & 0.23 &   4 &    $-0.47$ & 0.20 & 0.24 &   4 \\
LaII &$-1.38 $ & 0.08 & 0.09 &   4 &    $-1.27$ & 0.05 & 0.08 &   4 &    $-1.37$ & 0.10 & 0.17 &   1 \\
CeII &   $-1.15$ & 0.06 & 0.11 &   4 &    $-0.78$ & 0.06 & 0.10 &   4 &    \nodata & \nodata & \nodata & \nodata \\
PrII &$<$$ -1.39$ & \nodata & \nodata & \nodata &    \nodata & \nodata & \nodata & \nodata &    \nodata & \nodata & \nodata & \nodata \\
NdII &$-0.96 $ & 0.19 & 0.11 &  12 & $-0.60 $ & 0.23 & 0.12 &  15 & $-0.54$ & 0.20 & 0.24 &   1 \\
SmII &$-1.30$ & 0.17 & 0.12 &   9 & $-0.92$ & 0.12 & 0.11 &   6 &    \nodata & \nodata & \nodata & \nodata \\
EuII &$ -1.56$ & 0.06 & 0.08 &   3 & $-1.28 $ & 0.01 & 0.05 &   2 &    \nodata & \nodata & \nodata & \nodata \\
GdII &$-1.06$ & 0.24 & 0.15 &   3 &    \nodata & \nodata & \nodata & \nodata &    \nodata & \nodata & \nodata & \nodata \\
TbII &$<$$ -1.66$ & \nodata & \nodata & \nodata & $<$$-1.07$ & \nodata & \nodata & \nodata & $<$$-0.54$ & \nodata & \nodata & \nodata \\
DyII &$ -1.12$ & 0.33 & 0.10 &  13 & $ -0.58$ & 0.20 & 0.22 &   1 &    \nodata & \nodata & \nodata & \nodata \\
HoII &$<$$ -1.11$ & \nodata & \nodata & \nodata &    \nodata & \nodata & \nodata & \nodata &    \nodata & \nodata & \nodata & \nodata \\
ErII &$ -1.15$ & 0.02 & 0.06 &   3 &    \nodata & \nodata & \nodata & \nodata &    \nodata & \nodata & \nodata & \nodata \\
TmII &$<$$ -1.84$ & \nodata & \nodata & \nodata &    \nodata & \nodata & \nodata & \nodata &    \nodata & \nodata & \nodata & \nodata \\
YbII &$ -1.33$ & 0.20 & 0.27 &   1 &    \nodata & \nodata & \nodata & \nodata &    \nodata & \nodata & \nodata & \nodata \\
HfII &$<$$ -0.73$ & \nodata & \nodata & \nodata &    \nodata & \nodata & \nodata & \nodata &    \nodata & \nodata & \nodata & \nodata \\
OsI  &$<$$ -0.56$ & \nodata & \nodata & \nodata &    \nodata & \nodata & \nodata & \nodata & $<$$ 0.16$ & \nodata & \nodata & \nodata \\
\enddata
\end{deluxetable}

\begin{deluxetable}{lrcccrcccrccc}
\tablenum{8E}
\tablewidth{0pt}
\tablecaption{Abundances}
\tablehead{\colhead{Element} &\multicolumn{4}{c}{BD -18 5550} & 
\multicolumn{4}{c}{BD -17 6036} & \multicolumn{4}{c}{BD -11 145}    \\
\colhead{} & \multicolumn{1}{r}{log $\epsilon$} & \colhead{$\sigma$} 
& \colhead{$\sigma_{tot}$} & \colhead{N$_{lines}$} & \multicolumn{1}{r}{log $\epsilon$} 
& \colhead{$\sigma$} 
& \colhead{$\sigma_{tot}$} & \colhead{N$_{lines}$} & \multicolumn{1}{r}{log $\epsilon$} 
& \colhead{$\sigma$} 
& \colhead{$\sigma_{tot}$} & \colhead{N$_{lines}$} }
\startdata
NaI &$3.47 $ & 0.18 & 0.21 & 2 & $ 4.49$ & 0.10 & 0.22 & 1 & $ 4.08$ & 0.10 & 0.25 & 2 \\
MgI &$ 5.17$ & 0.11 & 0.14 & 8 & $ 5.54$ & 0.12 & 0.13 & 6 & $ 5.83$ & 0.04 & 0.13 & 5 \\
AlI & $2.94$ & 0.10 & 0.26 & 1 & $3.32$ & 0.10 & 0.24 & 1 & \nodata & \nodata & \nodata & \nodata \\
SiI &$5.01$ & 0.20 & 0.22 & 1 & $ 5.29$ & 0.20 & 0.23 & 1 & $ 5.52$ & 0.20 & 0.25 & 1 \\
CaI &$ 3.73$ & 0.07 & 0.04 & 15 & $ 4.01$ & 0.07 & 0.08 & 18 & $ 4.30$ & 0.08 & 0.09 & 18 \\
ScII & $0.00$ & 0.04 & 0.10 & 7 & $ 0.49$ & 0.12 & 0.14 & 8 & $ 0.64$ & 0.10 & 0.13 & 9 \\
TiI &$ 2.05$ & 0.07 & 0.08 & 25 & $ 2.42$ & 0.08 & 0.12 & 29 & $ 2.68$ & 0.10 & 0.13 & 20 \\
TiII &$ 2.09$ & 0.05 & 0.07 & 16 & $ 2.53$ & 0.08 & 0.12 & 18 & $ 2.72$ & 0.08 & 0.13 & 16 \\
VI & $0.88$ & 0.05 & 0.11 & 2 & $1.16$ & 0.06 & 0.13 & 3 & $1.32$ & 0.15 & 0.17 & 2 \\
VII & $0.88$ & 0.10 & 0.08 & 5 & $ 1.40$ & 0.08 & 0.10 & 2 & \nodata & \nodata & \nodata & \nodata \\
CrI & $2.43$ & 0.09 & 0.07 & 11 & $2.72$ & 0.10 & 0.12 & 12 & $2.96$ & 0.05 & 0.12 & 11 \\
CrII &$2.80$ & 0.09 & 0.11 & 2 & $ 3.11$ & 0.03 & 0.09 & 2 & $3.44$ & 0.04 & 0.08 & 2 \\
MnI & $2.03$ & 0.03 & 0.08 & 4 & $2.35$ & 0.07 & 0.09 & 3 & $2.70$ & 0.38 & 0.24 & 3 \\
MnII & $2.19$ & 0.05 & 0.16 & 3 & $2.37$ & 0.08 & 0.16 & 3 & \nodata & \nodata & \nodata & \nodata \\
FeI & $4.47$ & 0.10 & 0.08 & 151 & $4.75$ & 0.11 & 0.13 & 163 & $5.02$ & 0.11 & 0.15 & 135 \\
FeII & $4.46$ & 0.12 & 0.08 & 19 & $4.74$ & 0.08 & 0.10 & 19 & $5.04$ & 0.08 & 0.09 & 21 \\
CoI &$ 2.12$ & 0.12 & 0.12 & 7 & $ 2.40$ & 0.13 & 0.14 & 6 & $ 2.45$ & 0.20 & 0.18 & 3 \\
NiI &$ 3.27$ & 0.13 & 0.15 & 19 & $ 3.50$ & 0.13 & 0.17 & 22 & $3.70$ & 0.07 & 0.12 & 7 \\
ZnI &$ 1.94$ & 0.04 & 0.05 & 2 & $ 2.03$ & 0.04 & 0.07 & 1 & $ 2.35$ & 0.04 & 0.06 & 1 \\
YII & $-1.81$ & 0.03 & 0.05 & 6 & $-1.15$ & 0.08 & 0.11 & 9 & $-0.57$ & 0.06 & 0.11 & 4 \\
ZrII & $-1.22$ & 0.13 & 0.09 & 3 & $-0.48$ & 0.06 & 0.09 & 4 & $0.09$ & 0.11 & 0.14 & 1 \\
BaII & $-1.67$ & 0.15 & 0.16 & 3 & $-1.09$ & 0.15 & 0.18 & 4 & $ -0.29$ & 0.05 & 0.23 & 4 \\
LaII & $-2.43$ & 0.10 & 0.14 & 1 & $-1.86$ & 0.12 & 0.13 & 3 & $-1.39$ & 0.18 & 0.16 & 2 \\
CeII & \nodata & \nodata & \nodata & \nodata & \nodata & \nodata & \nodata & \nodata & $-0.96$ & 0.11 & 0.13 & 3 \\
PrII &$<$$ -1.69$ & \nodata & \nodata & \nodata & $<$$ -1.05$ & \nodata & \nodata & \nodata & \nodata & \nodata & \nodata & \nodata \\
NdII & \nodata & \nodata & \nodata & \nodata & $-1.31$ & 0.20 & 0.23 & 1 & $ -0.91$ & 0.20 & 0.17 & 2 \\
SmII & \nodata & \nodata & \nodata & \nodata & \nodata & \nodata & \nodata & \nodata & \nodata & \nodata & \nodata & \nodata \\
EuII & $-2.79$ & 0.20 & 0.22 & 1 & $ -2.21$ & 0.20 & 0.23 & 1 & $ -1.68$ & 0.20 & 0.22 & 1 \\
GdII & \nodata & \nodata & \nodata & \nodata & \nodata & \nodata & \nodata & \nodata & \nodata & \nodata & \nodata & \nodata \\
TbII &$<$$ -2.25$ & \nodata & \nodata & \nodata & $<$$ -1.61$ & \nodata & \nodata & \nodata & $<$$ -0.84$ & \nodata & \nodata & \nodata \\
DyII & $-2.28$ & 0.20 & 0.20 & 1 & $ -1.66$ & 0.20 & 0.22 & 1 & \nodata & \nodata & \nodata & \nodata \\
HoII &$<$$ -0.90$ & \nodata & \nodata & \nodata & $<$$ -0.76$ & \nodata & \nodata & \nodata & \nodata & \nodata & \nodata & \nodata \\
ErII & $-2.22$ & 0.10 & 0.12 & 1 & $ -1.84$ & 0.10 & 0.14 & 1 & \nodata & \nodata & \nodata & \nodata \\
TmII &$<$$ -1.47$ & \nodata & \nodata & \nodata & $<$$ -2.13$ & \nodata & \nodata & \nodata & \nodata & \nodata & \nodata & \nodata \\
YbII & $-2.79$ & 0.20 & 0.21 & 1 & $-1.94$ & 0.20 & 0.23 & 1 & \nodata & \nodata & \nodata & \nodata \\
HfII &$<$$ -1.02$ & \nodata & \nodata & \nodata & \nodata & \nodata & \nodata & \nodata & \nodata & \nodata & \nodata & \nodata \\
OsI &$<$$ -0.55$ & \nodata & \nodata & \nodata & $<$$ -0.41$ & \nodata & \nodata & \nodata & $<$$ -0.14$ & \nodata & \nodata & \nodata \\
\enddata
\end{deluxetable}

\begin{deluxetable}{lrcccrcccrccc}
\tablenum{8F}
\tablewidth{0pt}
\tablecaption{Abundances}
\tablehead{\colhead{Element} &\multicolumn{4}{c}{BD +4 2621} & 
\multicolumn{4}{c}{BD +5 3098} & \multicolumn{4}{c}{BD +8 2856} \\
\colhead{} & \multicolumn{1}{r}{log $\epsilon$} & \colhead{$\sigma$} 
& \colhead{$\sigma_{tot}$} & \colhead{N$_{lines}$} & \multicolumn{1}{r}{log $\epsilon$}
& \colhead{$\sigma$} 
& \colhead{$\sigma_{tot}$} & \colhead{N$_{lines}$} & \multicolumn{1}{r}{log $\epsilon$}
& \colhead{$\sigma$} 
& \colhead{$\sigma_{tot}$} & \colhead{N$_{lines}$} }
\startdata
NaI & \nodata & \nodata & \nodata & \nodata & \nodata & \nodata & \nodata & \nodata & $4.11$ & 0.10 & 0.28 & 2 \\
MgI &$ 5.49$ & 0.26 & 0.18 & 4 & $ 5.58$ & 0.17 & 0.14 & 6 & $ 6.11$ & 0.17 & 0.17 & 5 \\
AlI & $3.58$ & 0.10 & 0.26 & 1 & $ 3.74$ & 0.10 & 0.21 & 1 & $3.98$ & 0.10 & 0.27 & 1 \\
SiI &$ 5.69$ & 0.20 & 0.27 & 1 & $ 5.38$ & 0.20 & 0.24 & 1 & $ 5.99$ & 0.10 & 0.09 & 4 \\
CaI &$ 4.18$ & 0.06 & 0.09 & 5 & $ 4.12$ & 0.06 & 0.07 & 20 & $ 4.57$ & 0.09 & 0.12 & 21 \\
ScII &$ 0.71$ & 0.10 & 0.14 & 7 & $ 0.43$ & 0.08 & 0.14 & 8 & $ 1.10$ & 0.09 & 0.12 & 10 \\
TiI &$ 2.64$ & 0.09 & 0.14 & 20 & $ 2.47$ & 0.10 & 0.12 & 30 & $ 3.01$ & 0.16 & 0.19 & 38 \\
TiII &$ 2.83$ & 0.07 & 0.13 & 17 & $ 2.50$ & 0.06 & 0.13 & 17 & $ 3.20$ & 0.07 & 0.12 & 19 \\
VI & $1.39$ & 0.09 & 0.15 & 5 & $1.21$ & 0.07 & 0.13 & 3 & $1.74$ & 0.07 & 0.18 & 5 \\
VII &$ 1.68$ & 0.08 & 0.10 & 2 & $1.20$ & 0.13 & 0.11 & 3 & $ 1.97$ & 0.01 & 0.04 & 2 \\
CrI & $3.01$ & 0.05 & 0.13 & 9 & $2.78$ & 0.10 & 0.12 & 11 & $3.39$ & 0.06 & 0.17 & 14 \\
CrII &$ 3.45$ & 0.01 & 0.07 & 2 & $ 3.35$ & 0.15 & 0.13 & 2 & $ 3.82$ & 0.06 & 0.09 & 2 \\
MnI & $2.60$ & 0.05 & 0.10 & 4 & $2.31$ & 0.04 & 0.09 & 4 & $3.00$ & 0.16 & 0.13 & 6 \\
MnII &$ 2.89$ & 0.09 & 0.21 & 3 & $2.20$ & 0.21 & 0.18 & 2 & $3.22$ & 0.17 & 0.26 & 2 \\
FeI & $5.00$ & 0.15 & 0.17 & 69 & $4.78$ & 0.11 & 0.14 & 159 & $5.40$ & 0.17 & 0.19 & 166 \\
FeII & $4.99$ & 0.10 & 0.09 & 18 & $4.79$ & 0.10 & 0.10 & 20 & $5.41$ & 0.15 & 0.11 & 23 \\
CoI &$ 2.53$ & 0.13 & 0.17 & 5 & $ 2.33$ & 0.12 & 0.15 & 6 & $ 2.92$ & 0.18 & 0.24 & 4 \\
NiI &$ 3.78$ & 0.20 & 0.22 & 14 & $ 3.52$ & 0.15 & 0.17 & 21 & $4.04$ & 0.23 & 0.17 & 29 \\
ZnI & \nodata & \nodata & \nodata & \nodata & $ 2.10$ & 0.04 & 0.07 & 1 & $2.59$ & 0.04 & 0.05 & 2 \\
YII & $-0.69$ & 0.08 & 0.14 & 10 & $-0.87$ & 0.10 & 0.13 & 10 & $-0.15$ & 0.09 & 0.15 & 13 \\
ZrII & $0.03$ & 0.09 & 0.08 & 9 & $ -0.14$ & 0.19 & 0.14 & 4 & $ 0.57$ & 0.17 & 0.09 & 6 \\
BaII & $-1.21$ & 0.10 & 0.23 & 1 & $-0.96$ & 0.11 & 0.18 & 4 & $-0.07$ & 0.08 & 0.24 & 4 \\
LaII & $-2.27$ & 0.11 & 0.13 & 3 & $-1.63$ & 0.10 & 0.16 & 1 & $-0.96$ & 0.04 & 0.09 & 3 \\
CeII & \nodata & \nodata & \nodata & \nodata & \nodata & \nodata & \nodata & \nodata & $-0.65$ & 0.12 & 0.11 & 5 \\
PrII &$<$$ -1.30$ & \nodata & \nodata & \nodata & $<$$-1.03$ & \nodata & \nodata & \nodata & $-1.61$ & 0.14 & 0.13 & 2 \\
NdII & $-1.45$ & 0.20 & 0.18 & 2 & $ -1.13$ & 0.20 & 0.19 & 2 & $ -0.45$ & 0.24 & 0.12 & 10 \\
SmII & \nodata & \nodata & \nodata & \nodata & \nodata & \nodata & \nodata & \nodata & $ -0.90$ & 0.19 & 0.11 & 12 \\
EuII & $-2.61$ & 0.20 & 0.23 & 1 & $ -1.98$ & 0.07 & 0.12 & 2 & $ -1.16$ & 0.04 & 0.06 & 3 \\
GdII & \nodata & \nodata & \nodata & \nodata & \nodata & \nodata & \nodata & \nodata & $ -0.85$ & 0.20 & 0.22 & 1 \\
TbII &$<$$ -1.86$ & \nodata & \nodata & \nodata & $<$$ -1.49$ & \nodata & \nodata & \nodata & $-1.84$ & 0.15 & 0.17 & 1 \\
DyII & \nodata & \nodata & \nodata & \nodata & $-1.66$ & 0.20 & 0.22 & 1 & $-0.46$ & 0.26 & 0.12 & 7 \\
HoII &$<$$ -2.01$ & \nodata & \nodata & \nodata & \nodata & \nodata & \nodata & \nodata & $<$$ -0.72$ & \nodata & \nodata & \nodata \\
ErII & $-2.09$ & 0.10 & 0.14 & 1 & $ -1.60$ & 0.07 & 0.12 & 3 & $ -0.81$ & 0.01 & 0.09 & 3 \\
TmII &$<$$ -1.88$ & \nodata & \nodata & \nodata & $<$$ -1.81$ & \nodata & \nodata & \nodata & $ -1.71$ & 0.03 & 0.06 & 3 \\
YbII & $-2.49$ & 0.20 & 0.22 & 1 & $ -1.63$ & 0.20 & 0.25 & 1 & $ -0.82$ & 0.20 & 0.30 & 1 \\
HfII & \nodata & \nodata & \nodata & \nodata & $<$$ -0.36$ & \nodata & \nodata & \nodata & $<$$ 0.16$ & \nodata & \nodata & \nodata \\
OsI &$<$$ -0.56$ & \nodata & \nodata & \nodata & $<$$ 0.21$ & \nodata & \nodata & \nodata & $<$$ -0.07$ & \nodata & \nodata & \nodata \\
\enddata
\end{deluxetable}

\begin{deluxetable}{lrcccrcccrccc}
\tablenum{8G}
\tablewidth{0pt}
\tablecaption{Abundances}
\tablehead{\colhead{Element} &\multicolumn{4}{c}{BD +9 3223} & 
\multicolumn{4}{c}{BD +10 2495} & \multicolumn{4}{c}{BD +17 3248} \\
\colhead{} & \multicolumn{1}{r}{log $\epsilon$} & \colhead{$\sigma$} 
& \colhead{$\sigma_{tot}$} & \colhead{N$_{lines}$} & \multicolumn{1}{r}{log $\epsilon$} 
& \colhead{$\sigma$} 
& \colhead{$\sigma_{tot}$} & \colhead{N$_{lines}$} & \multicolumn{1}{r}{log $\epsilon$} 
& \colhead{$\sigma$} 
& \colhead{$\sigma_{tot}$} & \colhead{N$_{lines}$} }
\startdata
NaI &$ 4.71$ & 0.23 & 0.24 & 2 & $4.32$ & 0.10 & 0.18 & 2 & $ 4.76$ & 0.18 & 0.21 & 2 \\
MgI &$ 5.91$ & 0.07 & 0.10 & 5 & $6.01$ & 0.09 & 0.13 & 5 & $ 6.10$ & 0.05 & 0.11 & 5 \\
AlI & $3.53$ & 0.10 & 0.23 & 1 & $3.66$ & 0.10 & 0.23 & 1 & $3.71$ & 0.10 & 0.25 & 1 \\
SiI &$ 5.54$ & 0.20 & 0.22 & 1 & $5.72$ & 0.20 & 0.26 & 1 & $ 5.72$ & 0.20 & 0.23 & 1 \\
CaI &$ 4.51$ & 0.07 & 0.06 & 20 & $4.61$ & 0.07 & 0.10 & 22 & $4.65$ & 0.08 & 0.08 & 21 \\
ScII &$ 0.88$ & 0.05 & 0.14 & 8 & $1.06$ & 0.08 & 0.16 & 8 & $ 1.17$ & 0.08 & 0.16 & 10 \\
TiI &$ 3.01$ & 0.10 & 0.11 & 13 & $ 3.03$ & 0.10 & 0.14 & 30 & $3.13$ & 0.09 & 0.11 & 19 \\
TiII &$ 2.99$ & 0.06 & 0.15 & 16 & $ 3.13$ & 0.06 & 0.16 & 15 & $ 3.20$ & 0.05 & 0.16 & 16 \\
VI & $1.69$ & 0.05 & 0.12 & 1 & $1.82$ & 0.11 & 0.16 & 2 & $1.79$ & 0.05 & 0.12 & 3 \\
VII &$1.80$ & 0.10 & 0.16 & 1 & $ 1.94$ & 0.10 & 0.16 & 1 & $ 2.10$ & 0.10 & 0.17 & 1 \\
CrI & $3.33$ & 0.06 & 0.10 & 9 & $3.46$ & 0.04 & 0.14 & 14 & $3.49$ & 0.10 & 0.11 & 12 \\
CrII &$ 3.57$ & 0.04 & 0.10 & 2 & $3.85$ & 0.15 & 0.14 & 2 & $ 3.79$ & 0.05 & 0.11 & 2 \\
MnI & $2.87$ & 0.08 & 0.11 & 1 & $2.90$ & 0.07 & 0.12 & 3 & $3.01$ & 0.09 & 0.10 & 4 \\
MnII & \nodata & \nodata & \nodata & \nodata & \nodata & \nodata & \nodata & \nodata & \nodata & \nodata & \nodata & \nodata \\
FeI & $5.23$ & 0.09 & 0.11 & 128 & $5.44$ & 0.13 & 0.17 & 157 & $5.41$ & 0.12 & 0.13 & 139 \\
FeII & $5.24$ & 0.14 & 0.13 & 19 & $5.44$ & 0.11 & 0.13 & 23 & $5.41$ & 0.09 & 0.14 & 22 \\
CoI &$ 2.73$ & 0.04 & 0.12 & 2 & $ 3.04$ & 0.15 & 0.17 & 5 & $ 2.96$ & 0.33 & 0.21 & 4 \\
NiI &$ 3.99$ & 0.03 & 0.08 & 4 & $4.08$ & 0.14 & 0.11 & 19 & $ 4.19$ & 0.19 & 0.13 & 8 \\
ZnI &$ 2.45$ & 0.04 & 0.09 & 1 & $ 2.56$ & 0.04 & 0.08 & 2 & $ 2.55$ & 0.04 & 0.09 & 1 \\
YII & $-0.23$ & 0.09 & 0.14 & 3 & $-0.16$ & 0.08 & 0.14 & 6 & $0.10$ & 0.15 & 0.16 & 5 \\
ZrII &$ 0.44$ & 0.11 & 0.17 & 1 & $0.48$ & 0.11 & 0.17 & 1 & $ 0.75$ & 0.11 & 0.18 & 1 \\
BaII &$ -0.13$ & 0.04 & 0.19 & 4 & $0.03$ & 0.04 & 0.24 & 4 & $ 0.51$ & 0.09 & 0.26 & 4 \\
LaII & $-1.09$ & 0.04 & 0.14 & 2 & $-0.98$ & 0.06 & 0.13 & 4 & $ -0.53$ & 0.06 & 0.14 & 3 \\
CeII & \nodata & \nodata & \nodata & \nodata & $-0.70$ & 0.03 & 0.14 & 2 & $ -0.21$ & 0.01 & 0.14 & 2 \\
PrII & \nodata & \nodata & \nodata & \nodata & \nodata & \nodata & \nodata & \nodata & \nodata & \nodata & \nodata & \nodata \\
NdII &$ -0.34$ & 0.20 & 0.24 & 1 & $ -0.35$ & 0.20 & 0.18 & 3 & $ -0.06$ & 0.23 & 0.16 & 8 \\
SmII & \nodata & \nodata & \nodata & \nodata & \nodata & \nodata & \nodata & \nodata & $ -0.26$ & 0.15 & 0.16 & 4 \\
EuII &$ -1.63$ & 0.20 & 0.24 & 1 & $ -1.32$ & 0.20 & 0.24 & 1 & $ -0.80$ & 0.20 & 0.24 & 1 \\
GdII & \nodata & \nodata & \nodata & \nodata & \nodata & \nodata & \nodata & \nodata & \nodata & \nodata & \nodata & \nodata \\
TbII &$<$$ -0.64$ & \nodata & \nodata & \nodata & $<$$ -0.93$ & \nodata & \nodata & \nodata & $<$$ 1.32$ & \nodata & \nodata & \nodata \\
DyII & \nodata & \nodata & \nodata & \nodata & \nodata & \nodata & \nodata & \nodata & $ -0.13$ & 0.20 & 0.20 & 2 \\
HoII & \nodata & \nodata & \nodata & \nodata & \nodata & \nodata & \nodata & \nodata & \nodata & \nodata & \nodata & \nodata \\
ErII & \nodata & \nodata & \nodata & \nodata & \nodata & \nodata & \nodata & \nodata & \nodata & \nodata & \nodata & \nodata \\
TmII & \nodata & \nodata & \nodata & \nodata & \nodata & \nodata & \nodata & \nodata & \nodata & \nodata & \nodata & \nodata \\
YbII & \nodata & \nodata & \nodata & \nodata & \nodata & \nodata & \nodata & \nodata & \nodata & \nodata & \nodata & \nodata \\
HfII & \nodata & \nodata & \nodata & \nodata & \nodata & \nodata & \nodata & \nodata & \nodata & \nodata & \nodata & \nodata \\
OsI &$<$$ 0.66$ & \nodata & \nodata & \nodata & $<$$ 0.27$ & \nodata & \nodata & \nodata & $<$$ 0.84$ & \nodata & \nodata & \nodata \\
\enddata
\end{deluxetable}

\begin{deluxetable}{lccccccccc}
\tablenum{H}
\tablewidth{0pt}
\tablecaption{Abundances}
\tablehead{\colhead{Element} &\multicolumn{4}{c}{BD +18 2890} & 
\multicolumn{4}{c}{M92 VII-18} & \colhead{Sun} \\
\colhead{} & \multicolumn{1}{r}{log $\epsilon$} & \colhead{$\sigma$} 
& \colhead{$\sigma_{tot}$} & \colhead{N$_{lines}$} & \multicolumn{1}{r}{log $\epsilon$}
& \colhead{$\sigma$} 
& \colhead{$\sigma_{tot}$} & \colhead{N$_{lines}$} & \multicolumn{1}{r}{log $\epsilon$} }
\startdata
NaI &  $4.54$ & 0.10 & 0.19 &  2 &  \nodata & \nodata & \nodata & \nodata & 6.33\\
MgI &$ 6.44$ & 0.12 & 0.15 &  5 & $ 5.59$ & 0.08 & 0.22 &  1 & 7.58 \\
AlI &  $4.10$ & 0.10 & 0.21 &  1 &  \nodata & \nodata & \nodata & \nodata & 6.47 \\
SiI &$ 6.28$ & 0.20 & 0.13 &  4 &  \nodata & \nodata & \nodata & \nodata & 7.55\\
CaI &$ 5.06$ & 0.09 & 0.13 & 25 & $ 4.46$ & 0.11 & 0.30 &  5 & 6.36 \\
ScII &$ 1.51$ & 0.15 & 0.18 & 10 & $ 0.99$ & 0.01 & 0.13 &  2 & 310 \\
TiI &$ 3.40$ & 0.09 & 0.16 & 33 &  $2.56$ & 0.19 & 0.60 & 10 & 4.99 \\
TiII &$ 3.55$ & 0.13 & 0.16 & 18 & $ 3.12$ & 0.11 & 0.13 &  9 & 4.99 \\
VI  &  $2.22$ & 0.05 & 0.16 &  1 &  $1.26$ & 0.09 & 0.74 &  4  & 4.00\\
VII &$ 2.60$ & 0.10 & 0.16 &  1 &  \nodata & \nodata & \nodata & \nodata & 4.00\\
CrI &  $3.84$ & 0.08 & 0.17 & 14 &  $2.98$ & 0.07 & 0.56 &  7 & 5.67\\
CrII &$ 4.20$ & 0.03 & 0.10 &  2 & $ 3.63$ & 0.05 & 0.26 &  1 & 5.67 \\
MnI &  $3.36$ & 0.08 & 0.12 &  6 &  $2.76$ & 0.13 & 0.34 &  4 & 5.39\\
MnII &  \nodata & \nodata & \nodata & \nodata &  \nodata & \nodata & \nodata & \nodata & 5.39\\
FeI &  $5.79$ & 0.15 & 0.20 & 165 &  $5.23$ & 0.06 & 0.47 &  9 & 7.52\\
FeII &  $5.77$ & 0.13 & 0.16 & 21 &  $5.28$ & 0.15 & 0.36 &  7 & 7.52\\
CoI &$ 3.39$ & 0.54 & 0.34 &  4 & $ 2.66$ & 0.23 & 0.47 &  3 & 4.92\\
NiI &  $4.50$ & 0.17 & 0.13 & 34 &  $3.92$ & 0.17 & 0.21 &  5 & 6.25\\
ZnI &$ 2.89$ & 0.04 & 0.09 &  2 &  \nodata & \nodata & \nodata & \nodata & 4.60\\
YII &  $0.38$ & 0.13 & 0.16 &  5 &  $-0.20$ & 0.08 & 0.13 &  4 & 2.24\\
ZrII &$ 1.03$ & 0.03 & 0.14 &  3 & $ 0.57$ & 0.06 & 0.10 &  3 & 2.60 \\
BaII &$ 0.63$ & 0.17 & 0.29 &  4 &  $-0.52$ & 0.10 & 0.30 &  1 & 2.13\\
LaII &$ -0.42$ & 0.05 & 0.15 &  4 &  $-1.25$ & 0.07 & 0.05 &  4 & 1.22\\
CeII &$ -0.08$ & 0.20 & 0.17 &  5 &  $-1.16$ & 0.09 & 0.11 &  3 & 1.55\\
PrII &  \nodata & \nodata & \nodata & \nodata &  \nodata & \nodata & \nodata & \nodata & 0.71\\
NdII &$ 0.12$ & 0.20 & 0.16 & 10 &  $-0.79$ & 0.25 & 0.12 &  5 & 1.50\\
SmII &$ -0.20$ & 0.13 & 0.15 &  5 & $ -1.16$ & 0.20 & 0.16 &  3 & 1.00\\
EuII &$ -0.80$ & 0.20 & 0.24 &  1 & $ -1.45$ & 0.09 & 0.07 &  2 & 0.51\\
GdII &  \nodata & \nodata & \nodata & \nodata & $-0.70$ & 0.20 & 0.22 &  1 &1.12\\
TbII &$<$$ -0.59$ & \nodata & \nodata & \nodata &  \nodata & \nodata & \nodata & \nodata & 0.33\\
DyII &$ -0.15$ & 0.20 & 0.18 &  3 & $ -1.12$ & 0.20 & 0.22 &  1 & 1.10\\
HoII &  \nodata & \nodata & \nodata & \nodata &  \nodata & \nodata & \nodata & \nodata & 0.50\\
ErII &  \nodata & \nodata & \nodata & \nodata & $ -1.00$ & 0.20 & 0.18 &  2 & 0.93\\
TmII &  \nodata & \nodata & \nodata & \nodata &  \nodata & \nodata & \nodata & \nodata & 0.13\\
YbII &  \nodata & \nodata & \nodata & \nodata &  $-1.24$ & 0.20 & 0.24 &  1 & 1.08\\
HfII &  \nodata & \nodata & \nodata & \nodata &  \nodata & \nodata & \nodata & \nodata & 0.88\\
OsI &$<$$0.61$ & \nodata & \nodata & \nodata &  \nodata & \nodata & \nodata & \nodata & 1.45\\
\enddata
\end{deluxetable}

\subsection{Hyperfine Splitting}
Elements with large contributions from isotopes with nuclear spins can
have appreciable HFS.
This affects the derived abundances
by desaturating the line, requiring a lower abundance to match the EW.
If the hyperfine constants $A$ and $B$ are known, then the energy
difference between the hyperfine levels can be calculated.
Table 9 lists the $A$ and $B$ constants for 
all the
levels involved in our transitions that had literature values.
The relative probabilities for hyperfine transitions were taken from 
the tables of
White \& Eliason (1933).  We could not find $A$ and $B$ values for some 
\ion{Mn}{1} lines.  For these we used the splittings derived by Booth \etal{} (1984)
based on a high-resolution study of the structure of \ion{Mn}{1} lines.
For the Ba
isotopes, we assumed the solar-system r-process fractions from Sneden \etal{} (1996).

Unfortunately,
although the situation has improved recently, there remain many levels that
have no measured hyperfine constants.  Not all lines that have measured EWs 
could be used to give an accurate abundance.  However, if lines are
weak enough that they have not started to saturate, then hyperfine 
splitting can be safely ignored.  For \ion{V}{2}, 
we could not find $A$ values for both levels of any of the lines
we could measure.  For \ion{V}{1}, we found them only for the 4459.75 \AA{}  line, which
has a typical strength of 10 m\AA, where hyperfine splitting is negligible.
For \ion{V}{1} and \ion{V}{2}, however, we have lines with a variety of strengths.
There is no tendency for lines with larger EW to give larger abundances
than those with EW $<$ 20 m\AA{}, indicating that the
hyperfine splitting effects are small.  We have therefore accepted all
\ion{V}{1} and \ion{V}{2} lines with EW $\leq$ 50 m\AA.  

\begin{deluxetable}{lrrr|lrrr}
\tablenum{9}
\tablewidth{0pt}
\tablecaption{Hyperfine $A$ and $B$ Constants}
\tablehead{
\colhead{Level}  & \colhead{$A$} & \colhead{$B$} & \colhead{Ref} &
\colhead{Level}  & \colhead{$A$} & \colhead{$B$} & \colhead{Ref} \\ 
\colhead{} & \colhead{MHz} & \colhead{MHz} & \colhead{} & 
\colhead{} & \colhead{MHz} & \colhead{MHz} & \colhead{}
}
\startdata
\multicolumn{4}{c|}{NaI} & z$^8$P$_{\frac{9}{2}}$ 
\hfso &  456.28 &  47.97 & 8 \\
3s$^2$S$_{\frac{1}{2}}$  &  885.81 &   0.00 & 1 & a$^6$S$_{\frac{5}{2}}$  
 &  $-$72.42 & 0.00 & 9  \\
3p$^2$P$_{\frac{1}{2}}$\hfso &   94.42 &   0.00 & 1 & e$^6$S$_{\frac{5}{2}}$&  808.53 &   0.00  & 8 \\
3p$^2$P$_{\frac{3}{2}}$\hfso  &   18.69 &   0.00 & 1 & e$^8$S$_{\frac{7}{2
}}$ &  737.48 & 0.00 & 8  \\
\multicolumn{4}{c|}{AlI}    & \multicolumn{4}{c}{MnII} \\
3s$^2$3p$^2$P$_{\frac{3}{2}}$\hfso &   94.25 &   18.76 & 2 & a$^5$D$_1$       &  $-$59 &   $-$53 & 10 \\ 
3s$^2$4s$^2$S$_{\frac{1}{2}}$\hfso &  421.00 &   0.00 & 3 & a$^5$D$_3$       &  5.8 &   $-$71 & 10 \\  
\multicolumn{4}{c|}{ScII}   & z$^5$P$_1$\hfso       &  $-$737 & 9 & 10 \\ 
a$^1$D$_2$       &  128.20 &   $-$0.39 & 4 & z$^5$P$_2$\hfso       &  $-$310.7 &   $-$87 & 10 \\
b$^1$D$_2$       &  149.36 &    7.82 & 4 & \multicolumn{4}{|c}{
CoI}   \\
z$^1$D$_2$\hfso  &  215.70 &   0.18 & 4 & z$^2$D$_{\frac{3}{2}}$\hfso & 1377.83 & 14.99 & 11
  \\
z$^3$D$_2$\hfso  &  125.70 &   0.06 & 4 & a$^2$F$_{\frac{5}{2}}$   & 1108.62 & 
 $-$92.93 & 11  \\
z$^3$D$_3$\hfso  &  101.80 &   0.24 & 4 &a$^2$F$_{\frac{7}{2}}$    &  391.53 & 
$-$106.13 & 11  \\
a$^3$F$_2$       &  290.67 &  $-$10.50 & 4 & b$^4$F$_{\frac{9}{2}}$    &  828.80 & 
$-$118.72 & 11  \\
a$^3$F$_3$       &  113.67 &  $-$12.62 & 4 & y$^2$G$_{\frac{7}{2}}$\hfso        &  
740.50 &  $-$21.00 & 11  \\
a$^3$F$_4$       &   38.36 &  $-$16.50 & 4 & y$^2$G$_{\frac{9}{2}}$\hfso        &  
439.12 &   32.98 & 11  \\ 
z$^1$F$_3$\hfso  &  193.10 &   $-$0.65 & 4 & z$^2$G$_{\frac{7}{2}}$\hfso        &  
919.46 &   95.93 & 11  \\
z$^2$F$_2$\hfso  &  366.80 &   $-$0.40 & 4 & z$^2$G$_{\frac{9}{2}}$\hfso        &  
493.75 &  629.56 & 11  \\
z$^3$F$_3$\hfso  &  205.40 &   $-$0.70 & 4  & z$^4$G$_{\frac{11}{2}}$\hfso       &  
771.96 &  209.00 & 11  \\ 
z$^3$F$_4$\hfso  &  102.30 &   $-$0.84 & 4  & z$^2$F$_{\frac{5}{2}}$\hfso        & 1
040.37 &  $-$47.97 & 11  \\
a$^1$G$_4$       &  135.23 &  $-$63.44 & 4 & z$^4$F$_{\frac{9}{2}}$\hfso        &  
810.03 &  $-$47.97 & 11  \\
a$^3$P$_1$       & $-$107.50 &   12.30 & 4 & \multicolumn{4}{|c}{$^{135}$BaII}   
\\
a$^3$P$_2$       &  $-$27.20 &   0.26 & 4 & 5d$^2$D$_{\frac{3}{2}}$   &  169.59 &
   28.95 & 12 \\
z$^3$P$_1$\hfso  &  255.00 &   0.10 & 4 & 5d$^2$D$_{\frac{5}{2}}$   &  $-$10.74 & 
  38.69  & 13 \\
z$^3$P$_2$\hfso  &  105.60 &   $-$0.21 & 4 & 6p$^2$P$_{\frac{1}{2}}$\hfso       & 
 664.60 &   0.00 & 14  \\
\multicolumn{4}{c|}{VI} & 6p$^2$P$_{\frac{3}{2}}$\hfso       & 
 113.00 &   59.00 & 14  \\
a$^6$D$_{\frac{7}{2}}$    &  382.23 &   0.00 & 5 & 6s$^2$S$_{\frac{1}{2}}$   & 3591.67 & 
  0.00  & 15 \\  
z$^6$P$_{\frac{5}{2}}$    &   29.08 &   0.00 & 6 & \multicolumn{4}{|c}{$^{137}$BaII} \\
\multicolumn{4}{c|}{MnI} & 5d$^2$D$_{\frac{3}{2}}$   &  189.73 &   44.54 &  12  
\\
z$^6$P$_{\frac{3}{2}}$\hfso        &  569.90 &   17.99 & 7 & 5d$^2$D$_{\frac{5}{2}}$   & 
 $-$12.03 &   59.53 & 13  \\ 
z$^6$P$_{\frac{5}{2}}$\hfso        &  465.27 &  $-$77.95 & 7 & 6p$^2$P$_{\frac{1}{2}}$\hfso       &  743.70 &   0.00 & 14  \\ 
z$^6$P$_{\frac{7}{2}}$\hfso &  428.40 &   65.95 & 7  & 6p$^2$P$_{\frac{3}{2}}$\hfso       &  127.20 &   92.5
0 & 14  \\
z$^8$P$_{\frac{7}{2}}$\hfso &  546.52 & $-$101.93 & 7 & 6s$^2$S$_{\frac{1}{2}}$   & 4018.87 & 0.00 & 15  \\

\enddata
\tablerefs{(1) J\"onsson \etal{} (1996) and references therein; (2) Lew 
(1949);
(3) Jiang, Lundberg, \& Svanberg (1982); (4) Villemoes \etal{} (1992) and references therein;
(5) Palmeri \etal{} (1997); (6) Palmeri \etal{} (1995); (7) Walther (1962);
(8) Brodzinski \etal{} (1988); (9) Woodgate \& Martin (1949); (10) Holt, 
Scholl \& Rosner (1999);
(11) Pickering (1996) and references therein; (12) van Hove \etal{} (1985); (13)
Silverans \etal{} (1986); (14) Villemoes \etal{}(1993); (15) Becker, Blatt, \& Werth (1981)} 
\end{deluxetable}

\subsection{Error Analysis}

As mentioned above, we estimate our 1-$\sigma$ errors to be $\pm 100 K$ 
for T$_{\rm eff}$, $\pm 0.3$ for
log g and $\pm 0.3$ km/s for $\xi$.  Errors in the abundances
caused by uncertainties in the metallicity of the model atmosphere were
small compared to other sources, and we will no longer consider 
$\delta$[Fe/H]$_{mod}$
in our error analysis.  To determine our random errors, we will follow in
general the treatment of McWilliam \etal{} (1995b).  The variance in the 
abundance
of an element can be written
{\setlength
\arraycolsep{2pt}
\begin{eqnarray}
\sigma^2_{\rm{log}\epsilon}= \sigma^2_{rand} +
\left({\partial \rm{log}\epsilon\over\partial T}\right)^2 
\sigma^2_T + 
\left({\partial \rm{log}\epsilon\over\partial \mlogg}\right)^2 
\sigma^2_{\mlogg}  +
\nonumber\\ 
\left({\partial \rm{log}\epsilon\over\partial \xi}\right)^2 \sigma^2_{\xi} + 
2\biggl[\left({\partial \rm{log}\epsilon\over\partial T}\right)
\left({\partial \rm{log}\epsilon
\over\partial \mlogg}\right)\sigma_{T\mlogg} + 
\nonumber\\
 \left({\partial 
\rm{log}\epsilon\over\partial \xi}\right) \left({\partial \rm{log}\epsilon\over\partial 
\mlogg}\right) \sigma_{\mlogg \xi} +
 \left({\partial \rm{log}\epsilon\over\partial \xi}\right)\left({\partial 
\rm{log}\epsilon\over T}\right) \sigma_{\xi T} \biggr] 
\end{eqnarray}}
$\sigma^2_{rand}$ is the error due to EW and log \gf{} errors.  If we had
multiple lines to determine the abundance, we adopted the standard
error of the mean for $\sigma_{rand}$.  For some elements, e.g. Nd, Sm,
and Si, we could measure more than four lines only in some stars.  
We adopted 
an representative 
standard error of the sample ($\sigma_{s,avg}$) from the dispersion
seen in those stars.  
Then, for stars
with EWs for fewer than four lines for a particular element, 
their $\sigma_{rand} = \sigma_{s,avg}/\sqrt{N_{lines}}$.  We determined the 
partial
derivatives,
$({\partial \rm{log}\epsilon / \partial T})$,
$({\partial \rm{log}\epsilon / \partial \mlogg})$, and
$({\partial \rm{log}\epsilon / \partial \xi})$ by varying the model 
atmosphere of each star $\pm 100 K$ in \teff{}, then $\pm 0.3$ in
\logg and finally $\pm 0.3$ km/s in $\xi$ and redetermining the abundance
of each element using each model atmosphere.  The partial derivatives were
the average of ($\Delta$ log $\epsilon$)/($\Delta$ parameter) from the plus and
minus case.

The covariances $\sigma_{T \mlogg}$, $\sigma_{\xi \mlogg}$, and 
$\sigma_{T \xi}$ measure the correlation between our determinations
of the atmospheric parameters.  We found $\sigma_{T \xi}$ to be negligible,
because there was not a strong correlation between equivalent width
and E.P. for our \ion{Fe}{1} lines.
To calculate $\sigma_{T \mlogg}$, we picked 
BD $-$17 6036 as a representative case and picked 20 \teff{} values drawn from
a Gaussian distribution with a mean at 4750 K and a $\sigma$ of 100K.
For each of these cases, we created a new model atmosphere and then
adjusted \logg to force agreement between \ion{Fe}{1} and \ion{Fe}{2}.  The covariance
is then

\begin{equation}
\sigma_{T \logg}=\frac{1}{N}\sum_{i=1}^N \left(T_i - \overline{T}\right)
\left(\mlogg_i - \overline{\mlogg}\right)
\end{equation}
The correlation coefficient is defined as
\begin{equation}
\rho \left(T,\mlogg\right)={\sigma_{T \mlogg}\over \sigma_{T} \sigma_{\mlogg}}
\end{equation}
and is 0 for independent variables and 1 for complete correlation.
For \teff{} and log {\it g}, $\sigma_{T \mlogg}$ is 13.0 and $\rho(T,\mlogg)$
is 0.46, which agrees with our qualitative impression that our \teff{}
and \logg estimates are highly correlated.  We performed a similar 
calculation to determine $\sigma_{\xi \mlogg}$, picking 20 $\xi$s and
varying \logg to get ionization balance.  These two variables are
less correlated with $\sigma_{\xi \mlogg}=-0.02$ and 
$\rho(\xi,\mlogg)=-0.10$.

The above analysis gives the error in [M/H].  However, we are usually
interested in the ratio of two elements we have measured.  In that case,
the error in [A/B] is 

\begin{equation}
\sigma^2(A/B)=\sigma^2(A)+\sigma^2(B)-2 \sigma_{A,B}
\end{equation}
where the covariance between two abundances is given by 
{\setlength
\arraycolsep{2pt}
\begin{eqnarray}
\sigma_{A,B} = \left({\Pd \ep _A\over\Pd T}\right)\left({\Pd \ep _{B}\over\Pd 
T}\right)\sigma^2_T  + \hspace{0.8in}  
\nonumber\\
\left({\Pd \ep _A\over\Pd \mlogg}\right)
\left({\Pd \ep _B\over\Pd \mlogg}\right) \sigma^2_{\mlogg}  +  
\left({\Pd \ep _A\over\Pd \xi}\right) \left({\Pd \ep _B\over\Pd \xi}\right)
\sigma^2_{\xi} 
\nonumber\\
+ \biggl[\left({\Pd \ep _{A}\over\Pd T}\right)\left({\Pd \ep _{B}\over\Pd 
\mlogg}\right)  +  \left({\Pd \ep _{A}\over\Pd \mlogg}\right)
\left({\Pd \ep _{B}\over\Pd T}\right)\biggr] \sigma_{T \mlogg} 
\nonumber\\ 
+ \biggl[\left({\Pd \ep _{A}\over\Pd \xi}\right)\left({\Pd \ep _{B}\over\Pd 
\mlogg}\right)  + 
\left({\Pd \ep _{A}\over\Pd \mlogg}\right)
\left({\Pd \ep _{B}\over\Pd \xi}\right)\biggr]\sigma_{\xi \mlogg}    
\end{eqnarray}}

The derived abundances of some pairs of elements, 
such as \ion{Y}{2} and \ion{Zr}{2} or \ion{Fe}{1} and \ion{Ni}{1} have
similar sensitivities to changes in model atmosphere parameters, 
so the error in their
ratios is close to adding their $\sigma_{rand}$ in quadrature.
Other ratios that we are interested in, such as [Y/Fe] and [Ba/Fe] are more
susceptible to changes in the model atmosphere parameters.
In Table 8, we have include both the $\sigma_{rand}$ and the total 
$\sigma_{\rm{log}\epsilon}$
for each element.  When analyzing these results, we can use equation (6)
with the derivatives and covariances found here to determine error bars
for any abundance ratio.

\begin{figure}[htb]
\begin{center}
\includegraphics[width=3.0in,angle=0]{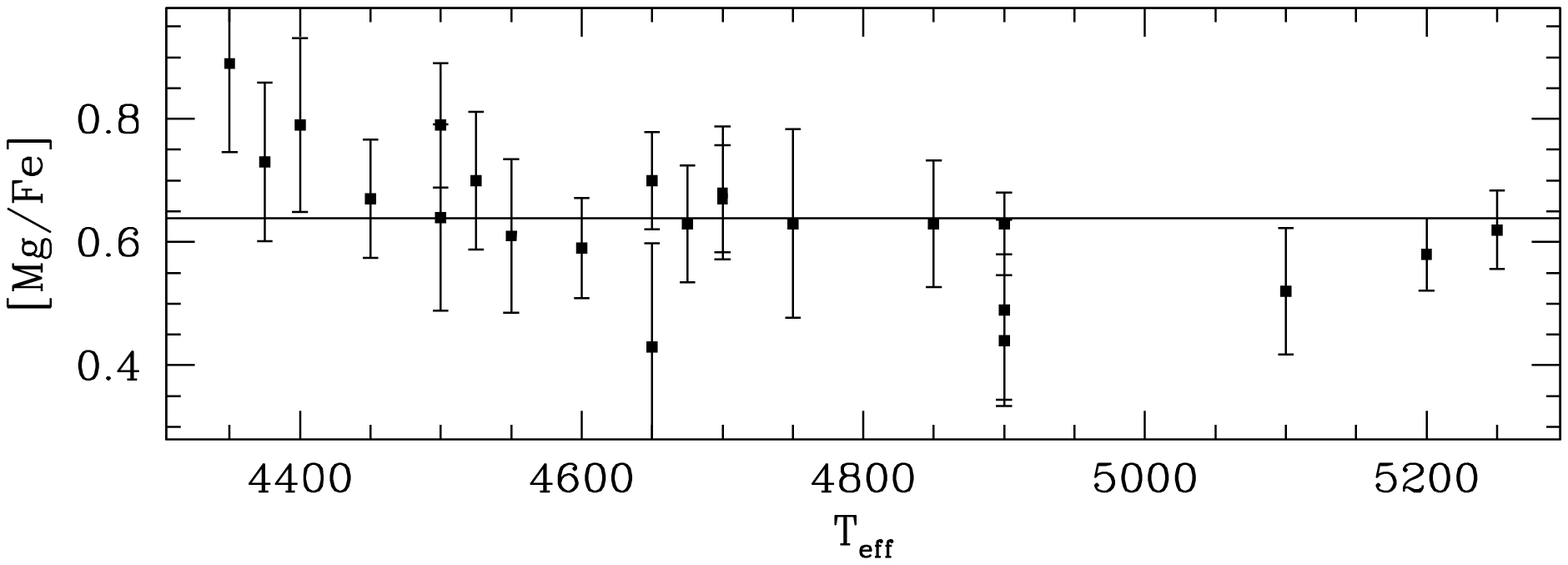}
\caption{[Mg/Fe] vs. \teff{} for the 22 field stars in our data.  The
slope apparent in this plot is due to the correlations of errors between
\teff and [Mg/Fe].}

\end{center}
\end{figure}

The powerful effect of correlated errors in \teff{} and
\logg{} on abundance ratios is illustrated by plotting \teff{} vs. 
[\ion{Mg}{1}/Fe] (Figure 9).  The lower the temperature, the
higher the [Mg/Fe] value is.  This would be of great concern, except
it is predicted by the reactions of Mg and Fe to changes in \teff{} and
the correlated change in log {\it g}.  If there is a random
error in \teff{} in the positive direction, the derived abundance
of both \ion{Mg}{1} and \ion{Fe}{1} increase, but Fe increases more,
leading to a decrease in [Mg/Fe].  This increase in \teff{} then
requires an increase in \logg to achieve ionization balance between
\ion{Fe}{1} and \ion{Fe}{2}.  Unfortunately, a change in \logg also
results in a decrease in the derived [Mg/Fe] ratio.  The result is 
correlation between \teff{} and [Mg/Fe].  To explore the magnitude
of this effect, we ran Monte Carlo tests.  The true [Mg/Fe] was assumed
to be constant.  We chose a \teff{} in our temperature range and
included a random error drawn from a Gaussian distribution 
with a $\sigma$ of 100 K.  $\Delta$\logg was calculated by assuming
a slope of 0.3dex/100K, derived from our BD $-$17 6036 tests.
The effect on [Mg/Fe] was determined using the 
($\Delta$ log $\epsilon$)/($\Delta$ parameter) values for the star closest
in temperature to our randomly selected temperature.  Using constant
values instead did not affect the analysis.  We calculated
Pearson's r statistic for 10,000 tests of 22 stars each and compared
the results with the r statistic for our observed sample.  
If there were no correlation between [Mg/Fe] and \teff{}, the r statistic
of our Monte Carlo tests should average zero.  Instead, its average is
$-$0.29, and our observed value of $-$0.56 is $ <$ a $2-\sigma$
result.  This conclusion was not changed by including additional scatter
in [Mg/Fe] because of $\sigma_{rand}$ or errors in $\xi$ or \logg{}
unassociated with our \teff{} error.  Even the effect of the   
the $\xi-$\logg correlation is small compared with the \teff$-$\logg one.
 If our $1-\sigma$ \teff{} errors
are 25 K instead of 100 K, then the [Mg/Fe] slope is significant,
but given the previous discussions, error bars that small are
unsupportable.  There is a tight relationship between our \teff{} and
\logg{} values and our \teff{} and $\xi$ values.  The former occurs both
because most of our stars are on the RGB and because of the correlation
of errors discussed above.   The rise in $\xi$ as the luminosity increases
has been noted by many authors (e.g. McWilliam \etal{} (1995b).)  Therefore
the correlation in errors between \teff{} and [Mg/Fe] leads to
slopes in the [Mg/Fe] vs. $\xi$, although the effect of errors in $\xi$
on the calculated abundances of Mg and Fe is small and similar.
In fact, trends with \teff{} are expected for most [M/Fe].  The error analysis
described in Equation (6) takes into consideration all of these correlations.  For
[Mg/Fe], for example, we find that the r.m.s. value for the 22 field stars is equal to the
1-sigma errors calculated from (6), so we would not conclude there was real scatter in
[Mg/Fe] despite a slope in the \teff-[Mg/Fe] plot.  This discussion does not
preclude the existence of other effects, such as systematically too low
\logg values at lower \teff.  But it clearly is first important to consider
the errors, as this explains most of the correlation found in our data.

\section{Summary}
We present the abundances of up to 30 elements in 23 metal-poor stars.
We have considered the effect on the abundances if we chose 
different model atmospheres.  In addition to the random errors in
our \teff{}, \logg{} and $\xi$ measurements, we examined several systematic
errors.
Among the alternatives we considered
were higher \logg{} models, 
higher \teff{} models, $\alpha$-enhanced models, models without overshooting,
MARCS models, as well as modified Kurucz models with depth-dependent
$\xi$ or a different temperature structure in the upper layers.
Our error analysis, adopted from McWilliam
\etal{} (1995b), takes into account correlations among the random
components of error in our model atmosphere determinations.  This can be very important
for some abundance ratios such as [Mg/Fe].  Errors
associated with systematic problems mentioned above have been tabulated
for three stars.  The use of $\alpha$-enhanced models did not change
the abundances by an appreciable amount.
In the other cases, the effect on the relative abundances, at least,
was usually small ($<$ 0.05 dex).  Further improvements (e.g. Asplund \etal{}
2000a,b) in
model atmosphere will likely result in resolution of several problems
discussed in this paper.  Having model-independent quantities, such
as bolometric corrections, radii, and distances, would also establish
what part of the analysis is at fault when discrepancies arise.
It is hoped that the next-generation Hipparcos missions will reach far
enough to determine the parameters for metal-poor giants.

\acknowledgements
I would like to thank Mike Bolte, under whose guidance as my thesis
advisor at UCSC most of this work was done.  Providing advice and support
as well were Chris Sneden, Graeme Smith, Bob Kraft, Ruth
Peterson  and Andy McWilliam.
Ruth Peterson also 
improved this paper with detailed comments on an initial draft.  My thanks
to the anonymous referee, who provided a careful and thoughtful 
review.
I acknowledge support from an NSF Graduate Student Fellowship and
a UCSC Dissertation Year Fellowship.  This work was completed with support
from a Carnegie Fellowship.  
Some of the data presented herein was obtained at the W.M. Keck
Observatory, which is operated as a scientific partnership among
the California Institute of Technology, the University of California
and the National Aeronautics and Space Administration.  The Observatory
was made possible by the generous financial support of the W.M. Keck
Foundation.

\clearpage

\end{document}